# A Conditional Point Cloud Diffusion Model for Deformable Liver Motion Tracking Via a Single Arbitrarily-Angled X-ray Projection

## Running Title:

## Deformable Liver Motion Tracking Via a Single X-ray Projection

Jiacheng Xie

Hua-Chieh Shao

Yunxiang Li

Shunyu Yan

Chenyang Shen

Jing Wang

You Zhang

*The Advanced Imaging and Informatics for Radiation Therapy (AIRT) Laboratory*
*The Medical Artificial Intelligence and Automation (MAIA) Laboratory*
*Department of Radiation Oncology, University of Texas Southwestern Medical Center, Dallas, TX 75390, USA*

Corresponding address:

You Zhang
Department of Radiation Oncology
University of Texas Southwestern Medical Center
2280 Inwood Road
Dallas, TX 75390
Email: You.Zhang@UTSouthwestern.edu
Tel: (214) 645-2699

## Abstract

*Objective.* Deformable liver motion tracking using a single X-ray projection enables real-time motion monitoring and treatment intervention. We introduce a conditional point cloud diffusion model-based framework for accurate and robust liver motion tracking from arbitrarily angled single X-ray projections. *Approach.* We propose a conditional point cloud diffusion model for liver motion tracking (PCD-Liver), which estimates volumetric liver motion by solving deformable vector fields (DVFs) of a prior liver surface point cloud, based on a single X-ray image. It is a patient-specific model of two main components: a rigid alignment model to estimate the liver's overall shifts, and a conditional point cloud diffusion model that further corrects for the liver surface's deformation. Conditioned on the motion-encoded features extracted from a single X-ray projection by a geometry-informed feature pooling layer, the diffusion model iteratively solves detailed liver surface DVFs in a projection angle-agnostic fashion. The liver surface motion solved by PCD-Liver is subsequently fed as the boundary condition into a UNet-based biomechanical model to infer the liver's internal motion to localize liver tumors. A dataset of 10 liver cancer patients was used for evaluation. We used the root mean square error (RMSE) and 95-percentile Hausdorff distance (HD95) metrics to examine the liver point cloud motion estimation accuracy, and the center-of-mass error (COME) to quantify the liver tumor localization error. *Main Results.* The mean (±s.d.) RMSE, HD95, and COME of the prior liver or tumor before motion estimation were 8.82 mm (±3.58 mm), 10.84 mm (±4.55 mm), and 9.72 mm (±4.34 mm), respectively. After PCD-Liver's motion estimation, the corresponding values were 3.63 mm (±1.88 mm), 4.29 mm (±1.75 mm), and 3.46 mm (±2.15 mm). Under highly noisy conditions, PCD-Liver maintained stable performance. *Significance.* This study presents an accurate and robust framework for liver deformable motion estimation and tumor localization for image-guided radiotherapy.

## 1. Introduction

In modern image-guided radiotherapy (IGRT), accurate localization of target and normal tissues is essential to maximize the therapeutic dose delivered to tumors while sparing adjacent healthy organs (Jaffray, 2012). This precision is crucial not only for treatment efficacy but also for minimizing toxicity risks. However, achieving accurate localization is particularly challenging for thoracic and abdominal tumors, such as those in the lung and the liver, where respiration-induced motion introduces substantial uncertainties (Shirato *et al.*, 2004; Yoganathan *et al.*, 2017). These tumors can move significantly throughout a single breathing cycle with substantial inter-cycle variations, introducing dose delivery discrepancies that compromise treatment outcomes (Remouchamps *et al.*, 2003). The risk of radiation exposure to surrounding or adjacent organs, such as lungs, liver, heart, kidneys, and other gastrointestinal organs, is also heightened without proper motion management. Intra-treatment motion management is therefore a fundamental need in IGRT (Bertholet *et al.*, 2019). Among the proposed/developed motion management techniques, real-time motion tracking and plan adaptation offer an elegant end-to-end solution by proactively adjusting dose delivery based on time-resolved motion and anatomy variations. It allows free-breathing-based treatments, minimizing the physiological/psychological burdens on patients as well as the associated compliance challenges and reduced duty cycles. However, real-time motion tracking poses a high demand for real-time imaging, which is constrained by current technological limitations (Skouboe *et al.*, 2019; Lombardo *et al.*, 2024). Cone-beam computed tomography (CBCT), a commonly employed technique for on-board 3D volumetric imaging, offers essential anatomical details for tumor and normal tissue localization. However, the long acquisition time (~ 1 minute or longer) associated with traditional CBCT imaging is unable to track respiratory motion within the temporal constraints required for real-time plan adaptation, where updates (imaging + plan adaptation) are ideally under 500 milliseconds to address rapid anatomy variations (Zhang *et al.*, 2024).

Current approaches to real-time imaging and motion tracking often rely on two-dimensional (2D) X-ray projections due to their rapid imaging speed (Seppenwoolde *et al.*, 2002; Keall *et al.*, 2004). Relying on temporal intensity gradients to infer 2D motion between consecutive frames, optical flow algorithms have been employed to derive motion vectors from sequential X-ray projections to estimate lung tumor motion, with reported sub-millimeter accuracy in both static and rotating gantry settings (Teo *et al.*, 2013; Fontaine *et al.*, 2017). However, 2D imaging lacks the depth information required for comprehensive three-dimensional (3D) target tracking, leading to compromises in precision and robustness (Dhont *et al.*, 2020). This limitation is further exacerbated in scenarios where complex, 3D tumor motion occurs, necessitating more advanced imaging and tracking solutions. Emerging real-time imaging techniques seek to overcome these limitations by incorporating supplementary information, such as optical surface imaging, to enhance localization accuracy (Shao *et al.*, 2023b; Batista *et al.*, 2020). Surface imaging can track external anatomical motion and, in combination with single or multiple X-ray projections, infer internal organ motion. However, as noted in recent studies, this approach requires a strong correlation between external surface movement and internal tumor motion, which may vary among patients and tumor locations (Freislederer *et al.*, 2020; Li, 2022). Additionally, surface imaging requires a clear line of sight to the patient's anatomy, which is not always feasible due to positioning constraints, or obstructions from linear accelerators/ancillary equipment (Al-Hallaq *et al.*, 2022).

Recent advancements in deep learning (DL) have introduced promising solutions for anatomical motion estimation. DL-based frameworks for real-time motion tracking leverage neural networks, which allow fast one-pass predictions without the iterative processes that most traditional algorithms require. These DL methods generally fall into two main categories: image reconstruction (Wang *et al.*, 2020) and image registration (Zou *et al.*, 2022). In the reconstruction approach, networks aim to generate 3D volumetric images directly from one or a few 2D X-ray projections (Shen *et al.*, 2019). Direct reconstructions offer detailed anatomical spatial mapping but faces inherent challenges due to extreme under-sampling, especially for low-contrast organs like the liver, where fine details are harder to reconstruct accurately (Tong *et al.*, 2020a). Additionally, reconstructed images typically require further segmentation or registration to localize tumors, adding complexity and potential sources of error, especially in anatomies with low X-ray contrast. Alternatively, registration-based techniques can directly propagate tumor contours from prior reference scans, allowing tumor tracking without full 3D reconstructions through correlating 3D anatomical motion with features extracted from 2D X-ray projections. Prior to the deep learning methods, an early machine learning-driven study, the motion modeling and free-form deformation (MM-FD) method, integrated principal component analysis (PCA)-based motion modeling with free-form deformation vector field (DVF) optimization to align prior CT images with 2D projections through 2D-3D deformation (Zhang *et al.*, 2013). Later developments on DL-based approaches have demonstrated promising accuracy and real-time performance in lung motion estimation. Foote et. al. developed a real-time 2D-3D deformable registration framework leveraging a DenseNet model trained on digitally reconstructed radiographs (DRRs) simulated from 4D-CTs to capture principal components of lung deformations during breathing, achieving over 1000 images per second throughput and deformation errors below 2 mm (Foote *et al.*, 2019). Similarly, Wei et. al. (Wei *et al.*, 2019) proposed a CNN-based method for real-time 3D tumor localization from single X-ray projections with a PCA-based lung motion model to generate corresponding DVFs, incorporating intensity correction and data augmentation to enhance robustness. Voxelmap (Hindley *et al.*, 2023) introduces a patient-specific framework that maps 2D images to 3D DVFs, enabling volumetric imaging during lung radiotherapy with submillimeter accuracy and an inference time of 50 milliseconds. Despite these advancements, most existing methods are optimized for lung imaging, where high tissue-air contrast supports accurate deformation estimation. However, these techniques often struggle with low-contrast regions, such as the liver site, where subtle intensity differences hinder motion estimation accuracy. Furthermore, most approaches are not angle-agnostic and rely on models trained for specific beam geometries or projection angles, limiting their generalizability to diverse imaging setups.

To address these limitations, our recent study introduced a novel approach specifically designed for real-time liver imaging and volumetric motion tracking using angle-agnostic X-ray projections. We proposed an X360 framework, which employed graph neural networks (GNNs) to track deformable liver surface motion from a single, angle-agnostic X-ray projection (Shao *et al.*, 2023a). The X-ray imaging has also been combined with optical surface imaging to better estimate the liver surface motion, and biomechanical modeling was introduced to propagate estimated liver boundary motion inside the liver to localize tumors in 3D (Shao et al., 2023b). Nevertheless, some limitations remain in our previous approaches. The X360 model, while being angle-agnostic, showed inferior accuracy compared to angle-specific models due to the added complexity of training for varying projection geometries. Additionally, its performance was found to be sensitive to noise which can obscure motion-related image features. Although the integration of surface imaging improved liver motion tracking and tumor localization accuracy, it introduced challenges including reliance on the potentially-inconsistent correlation between external body surface motion and internal liver deformation, as well as the susceptibility to errors in surface imaging due to external obstructions (Batista *et al.*, 2020).

Advances in diffusion models have marked a pivotal shift in generating high-quality images, establishing them as a robust alternative to traditional generative models such as GANs (Dhariwal and Nichol, 2021). Diffusion models employ a unique forward-reverse process: in the forward phase, noise is incrementally added to data until it approximates Gaussian noise; while in the reverse phase, this noise is systematically reduced, enabling precise reconstruction of the original data distribution. This approach has demonstrated state-of-the-art performance, particularly with score-based diffusion models, which are highly effective in capturing intricate statistical distributions across extensive datasets (Song et al., 2021; Chung and Ye, 2022). Enlightened by the strong capability of diffusion models to capture robust and stable priors, we aim to leverage this model for liver motion estimation, addressing key limitations of our previous frameworks. In this study, we proposed a conditional point cloud diffusion model for liver motion estimation (PCD-Liver), which progressively refines noisy point clouds to reconstruct liver shapes by solving liver surface node movements based on derived features from a single, arbitrarily angled X-ray projection. Compared to the previous X360 framework, our new approach overcomes critical weaknesses by significantly improving the accuracy of liver motion estimation using X-ray imaging and enhancing the model's robustness to noise. Unlike X360, which predicts liver motion in a two-step process—first using a rigid module followed by a deformable module—the diffusion model captures liver motion progressively. This makes it particularly well-suited for challenging scenarios involving low-contrast liver regions and noisy imaging conditions, paving the way for more reliable and efficient motion estimation in clinical practice. By PCD-Liver, a point cloud representation of the liver surface is utilized. Point clouds offer a flexible and efficient way to describe the liver's shape and motion without the constraints of predefined connectivity, making them well-suited for capturing complex deformations while preserving topological information of a surface (Tong *et al.*, 2020b). Moreover, point clouds can be directly converted into surface meshes, which are essential for biomechanical modeling. By integrating point clouds with mesh-based finite element methods, biomechanical models can leverage the strengths of both representations—using point clouds for data-driven liver surface motion estimation and meshes for physics-informed intra-liver deformation modeling (Zhang *et al.*, 2019). This synergy enables a more accurate and computationally efficient approach to tracking liver motion and tumor displacement.

The proposed PCD-Liver framework performed progressive denoising steps, each conditioned on extracted X-ray features, to capture liver deformation accurately. In diffusion models, the generation process begins with a sample from a Gaussian noise distribution and proceeds through a sequence of denoising steps. Each step incrementally refines the sample by removing a portion of the noise. This iterative process gradually transforms the noisy input into a sample from the target data distribution. The denoising steps are executed sequentially, each corresponding to a specific timestep. In the PCD-Liver framework, starting from a randomly initialized Gaussian point cloud, each denoising step moves the point cloud closer to the desired liver boundary DVF. An inherent geometry-informed perceptual feature pooling layer of the model enhanced this process by aligning 3D liver surface nodes with 2D X-ray features based on projection angle-specific geometry, allowing the model to adapt flexibly across varying projection angles. In addition, we utilized a pre-trained rigid alignment model to predict the liver's overall shifts, based on which the diffusion model further corrects for local liver deformations. With solved liver surface DVFs serving as the boundary condition, applying a sequential deep learning-based biomechanical model propagates the surface DVFs to infer intra-liver volumetric motion fields, achieving volumetric liver tumor localization. To elevate the model's accuracy, a composite loss function is employed, combining denoising and centroid-based mean square error (MSE) losses, as well as regularization terms like Laplacian and deformation energy losses. The hybrid loss function helps the predicted DVFs to be smooth, realistic, and aligned with 'ground-truth' motion patterns. The model is trained on augmented CT data reflecting diverse respiratory motions, simulating different projection angles with additional noise for robustness enhancement. For evaluation, liver boundary and tumor localization accuracy are assessed using metrics like root mean square error (RMSE), 95-percentile Hausdorff distance (HD95), and center-of-mass error (COME). Results show that this approach significantly outperforms the benchmark X360 model in both accuracy and robustness, without requiring additional surface imaging.

## 2. Materials and Methods

### *2.1 Background of the conditional point cloud diffusion model*

**Diffusion Models.** Diffusion denoising probabilistic models are general-purpose generative models that operate by progressively adding noise to a data sample $\mathrm{X}_0 \sim \mathrm{q}(\mathrm{X}_0)$ drawn from a target distribution $\mathrm{q}(\mathrm{X}_0)$ via a series of steps. Each step (t) adds noise according to a predefined variance schedule modeled as $\mathrm{q}(\mathrm{X}_t|\mathrm{X}_{t-1}) = \mathcal{N}\left(\mathrm{X}_t; \sqrt{1-\beta_t}\mathrm{X}_{t-1}, \beta_t\mathbf{I}\right)$ where $\{\beta_t\}_{t=0}^{T}$are variance schedule hyper-parameters that control the diffusion rate of the process. This creates a sequence of noisy samples, with each $\mathrm{q}(\mathrm{X}_t|\mathrm{X}_{t-1})$ representing a Gaussian distribution. The $X_0$, $X_t$, and $X_{t-1}$ represent data samples (e.g., an image or a deformation/velocity field) drawn from the noise distribution at timestep 0, t, and t-1. In our PCD-Liver framework, $X_t$ is defined as the DVF required to map the prior liver surface to its target position. Using the reparameterization trick (Kingma and Welling, 2013), this process can be expressed as (Ho *et al.*, 2020b):

$$X_t \sim q(X_t|X_0); \quad X_t = \sqrt{\overline{\alpha_t}}X_0 + \epsilon\sqrt{1-\overline{\alpha_t}}, \tag{1}$$
$$\text{where } \alpha_t = 1-\beta_t,\ \overline{\alpha_t} = \prod_{s=0}^{t}\alpha_s, \text{ and } \epsilon \sim \mathcal{N}(0,\mathbf{I}).$$

A reverse diffusion process is performed to gradually remove noise from corrupted samples and recover the underlying data distribution. This process begins by drawing a sample $X_T$ from the noise distribution $q(X_T)$, which typically represents an isotropic Gaussian distribution. $X_T \sim q(X_T)$ denotes that $X_T$ is sampled from a high-variance prior distribution, serving as the starting point for the denoising process. The goal is to iteratively transform this noisy sample into one that follows the target distribution $q(X_0)$. This is achieved by approximating the conditional distribution $q(X_{t-1}|X_t)$ using a neural network $s_\theta(X_{t-1}|X_t) \approx q(X_{t-1}|X_t)$. The model gradually refines the sample by iteratively drawing from $q(X_{t-1}|X_t)$ until it reaches $q(X_0)$, which closely resembles real data. When the step size in this reverse process is sufficiently small, $q(X_{t-1}|X_t)$ can be well-approximated by an isotropic Gaussian with a fixed small covariance. Ultimately, the neural network $s_\theta$ is trained to predict and remove noise at each step, allowing the model to generate high-quality samples that match the original data distribution.

**Point Cloud Diffusion Models.** In this study we propose to model the structure of the liver surface via a point cloud. The movement of the point cloud points, correspondingly, will describe the motion/deformation of the liver surface. We consider a 3D point cloud with $N$ points as a 3N-dimensional object and train a diffusion model $s_\theta: R^{3N} \to R^{3N}$. This model progressively denoises the positions of points from an initial spherical Gaussian distribution to form a recognizable shape to describe the liver surface motion. At each step, the network predicts the offset of each point from its current location, iterating through multiple steps to generate a sample that matches the distribution $q(X_0)$ of the target liver surface motion. Similar to the standard diffusion models, the neural network is trained to predict the noise $\epsilon \in R^{3N}$ introduced in the most recent time step using an $L_2$ loss function (Zhou *et al.*, 2021), which minimizes the difference between the true noise and the predicted noise:

$$\mathcal{L} = E_{t\sim[1,T]} E_{\epsilon_t \sim \mathbb{N}(\mathbb{0},\mathbf{I})} [||\epsilon_t - s_\theta(X_t, t)||_2^2], \tag{2}$$

During inference, we start by sampling a random point cloud $X_T \sim \mathcal{N}(\mathbf{0}, \mathbf{I}_{3N})$ from a 3N-dimensional Gaussian distribution. The reverse diffusion process is then applied iteratively to produce the final sample $X_0$.

**Conditional Point Cloud Diffusion Models.** We formulate the 3D shape movement estimation as a conditional generation problem: the target distribution is the distribution $q(X_0 \mid C)$ conditioned on the input $C$. For the specific real-time liver motion estimation problem of this study, the conditional input $C$ will be features extracted from a single X-ray projection. In relevant computer vision studies that similarly estimate 3D objects from a single-view image, the conditions are mostly incorporated as learned global features of the single-view image (Choy *et al.*, 2016; Xie *et al.*, 2019; Xie *et al.*, 2020). However, this approach has a limitation in that it only enforces a weak geometric consistency between the input image and the estimated shape. Previous work (Melas-Kyriazi *et al.*, 2023) has shown that while this approach often produced plausible shapes, these shapes lacked geometric/location accuracy as they did not consistently align with the input image from the specified viewpoint. For real-time motion tracking and target localization, such accuracy is essential. In this study, we employed a geometry-informed perceptual feature pooling layer that associates each 3D liver surface point cloud node with extracted 2D features of each X-ray projection, based on the cone-beam imaging geometry. Such a feature pooling layer helps to effectively encode and preserve the geometry information into the conditions to generate real-time liver shapes consistent with the input x-ray imaging angle, allowing the model to work precisely and flexibly towards different input x-ray imaging angles (angle-agnostic).

### *2.2 Geometry-informed perceptual feature pooling layer*

We adopt the geometry-informed perceptual feature pooling layer from the work of X360 (Shao *et al.*, 2023a). Figure 1 illustrates the concept of this pooling layer. For X-ray projections at arbitrary angles, ResNet-50 was selected as the backbone for feature extraction, as it demonstrated superior performance compared to alternatives like VGG-16 and MobileNets (Shao *et al.*, 2022). After feature extraction, the image feature maps were then input into the pooling layer. The pooling layer aligns 3D liver surface nodes with 2D feature map points through a cone-beam projector, achieving geometry awareness by projecting each liver surface node using angle-specific cone-beam geometry onto 2D feature maps. For each liver surface node, its feature vector is pooled as bilinearly-interpolated feature values of the four neighboring feature pixels near its projected locations on all 2D feature maps. For each liver surface node, 3840 feature values are pooled, forming a comprehensive feature representation. The resulting feature vector, concatenated with the initial liver surface point cloud, is then inputted into neural networks for direct estimation of the liver surface DVF. Through this integration, our model adapts to projection angle-dependent features and enables angle-agnostic training and inference.

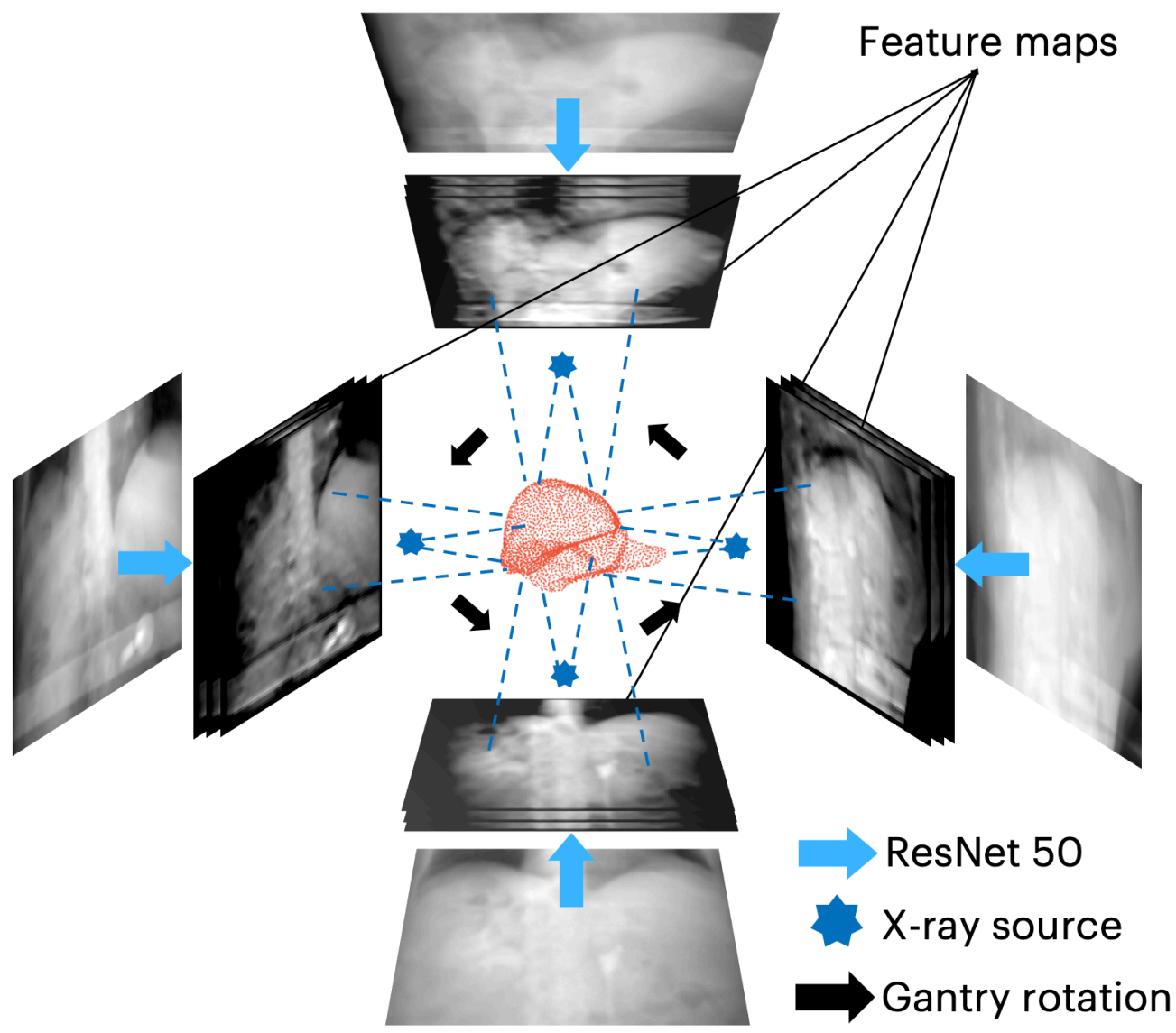

**Figure 1.** Illustration of the geometry-informed feature pooling layer. The geometry-informed perceptual feature pooling layer pools X-ray projection angle-dependent image features by mapping liver surface nodes onto feature maps extracted from ResNet-50. To enable angle-agnostic inference of liver boundary motion, the pooling layer integrates cone-beam geometry consistent with the X-ray projection, allowing extraction of relevant features specific to the current projection angle.

*2.3 The proposed PCD-Liver method*

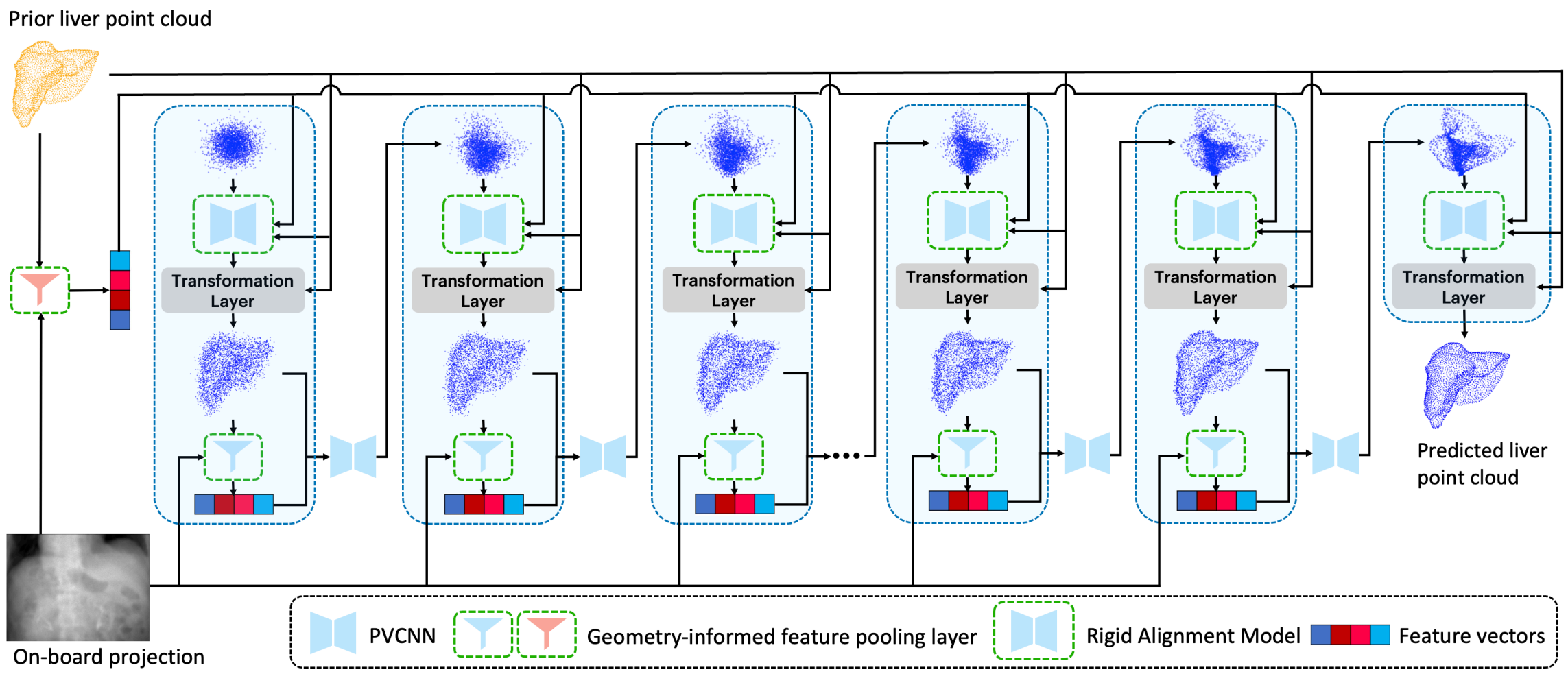

**Figure 2.** Overview of PCD-Liver. The geometry-informed feature pooling layers used by the rigid alignment model and the diffusion model are denoted as red and blue, respectively. Starting from Gaussian noise, in each step of the diffusion model (blue dashed box), the intermediate DVFs are repositioned by the pre-trained rigid alignment model and then applied to the prior liver surface point cloud to obtain an intermediate liver surface point cloud. Subsequently, in the feature pooling layer, the intermediate liver surface points are projected to a stack of 2D feature maps, yielding geometry-aware feature vectors. The feature maps are extracted from the onboard X-ray projection using a ResNet-50 that is simultaneously retrained in our network. The intermediate liver surface point cloud and feature vectors are fed into the diffusion model to produce denoised DVFs to input into the next step, forming an iterative loop. After 1000 steps, the predicted liver surface point cloud is obtained.

### *2.3.1 PCD-Liver workflow*

As aforementioned, we formulate the liver motion estimation as a conditional generation problem to predict the DVFs that capture the movement of the liver surface nodes, defined as:

$$\boldsymbol{X}_{DVF} = \boldsymbol{X}_{tar} - \boldsymbol{X}_{ref}, \tag{3}$$

where $\boldsymbol{X}_{ref}$ represents the initial 3D reference position of each liver surface node, and $\boldsymbol{X}_{tar}$ represents the target position of the node after deformation due to respiratory motion. The DVF at each node $\boldsymbol{X}_{DVF}$ captures the spatial displacement required to map the node to its target position, providing an essential representation of liver motion for subsequent analysis and treatment planning.

As shown in Fig. 2, the features of the on-board X-ray projection are extracted using a ResNet-50 that is jointly trained to generate a stack of 2D feature maps. These features are then processed by a geometry-informed feature pooling layer (color-coded red as in Fig. 2), which projects the prior reference liver surface point cloud onto the feature maps, producing feature vectors that contain node-specific motion information. The resulting feature vectors are concatenated with the coordinates of the prior reference liver surface point cloud and serve as input to a pre-trained rigid alignment model. During each denoising step, the intermediate DVFs are derived and repositioned by the rigid alignment model and then applied to the prior reference liver surface point cloud to obtain the intermediate liver surface point cloud. This intermediate surface point cloud is projected back onto the X-ray projection feature maps to obtain new feature vectors via another geometry-informed feature pooling layer (color-coded blue as in Fig. 2). Along with the intermediate liver surface point cloud, these updated feature vectors are then fed into conditional point cloud diffusion model $s_\theta$ for subsequent denoising steps, forming an iterative loop. Analytically, $s_\theta$ serves a function $R^{(3+F)N} \rightarrow$ $R^{3N}$ that predicts the noise $\epsilon$ from the augmented point cloud $\boldsymbol{X}_{DVF}^{+} = [\boldsymbol{X}_{DVF}, \boldsymbol{X}_t^{proj}]$. $F$ stands for the size of feature vector extracted for each liver surface node. The projected features $\boldsymbol{X}_t^{proj}$ are given by:

$$\boldsymbol{X}_t^{proj} = \mathcal{P}_{G_I}(I, \boldsymbol{X}_t) \qquad t \in [0, T], \tag{4}$$

where $I$ is the X-ray projection, $G_I$ is the corresponding gantry angle, $\mathcal{P}$ is the geometry-informed feature pooling layer (Fig. 1), and $\boldsymbol{X}_t$ denotes the intermediate point cloud where $\boldsymbol{X}_t = \boldsymbol{X}_{ref}$ when $t = T$, where $t$ is the number of diffusion steps counted in a backward order. Eventually, the predicted DVFs are applied to the prior reference liver surface point cloud to generate the predicted liver surface point cloud that captures liver motion, as shown in Fig. 2.

#### *2.3.2 Rigid alignment model*

Diffusion models are designed to learn the target data distribution by progressively denoising Gaussian noise added to a standard Gaussian prior distribution (Zhou *et al.*, 2021). This training process often involves normalizing data, leading the model to produce outputs centered near the origin (zero-mean). However, in practical applications, the desired outputs, such as DVFs representing point cloud transformations, may not be centered close to the origin, due to inherent rigid translations. Such translations are difficult to be captured by the point cloud diffusion model. This discrepancy necessitates a mechanism to adjust the diffusion model's output to account for the rigid translations. To address this, we introduce a rigid alignment model designed to adjust the diffusion model's output. This model estimates the rigid transformation, which can be added to the predicted DVFs by the point-cloud diffusion model to yield DVFs reflecting real-world scenarios including both rigid and deformable motion. This rigid alignment approach is akin to point cloud registration techniques (Zhang *et al.*, 2022), where the goal is to find the optimal rigid transformation that aligns two point sets. Coupled with the point cloud diffusion model, our rigid alignment model adjusts the predicted DVFs of the diffusion model to account for the overall shifts of the liver surface point cloud. Specifically, the rigid alignment model uses feature vectors extracted by the geometry-informed pooling layer (Fig. 1) based on the prior reference liver surface point cloud and the X-ray projection. The feature vectors, concatenated with the coordinates of the prior reference liver points, were used as inputs for the rigid alignment model. The model outputs a point cloud-wise DVF, which is then averaged to generate a rigid translation vector to represent the rigid motion. The rigid translation vector is added to DVFs generated by the point-cloud diffusion model, to yield the complete DVFs.

#### *2.3.3 Point-Voxel CNN*

For both the rigid alignment model and the backbone of the diffusion model, we utilized a Point-Voxel CNN (PVCNN) (Liu *et al.*, 2019) due to its demonstrated effectiveness in 3D generation (Zhou *et al.*, 2021). Unlike purely voxel-based or point-based methods, PVCNN effectively balances memory efficiency and computational speed by leveraging the advantages of both representations. Traditional voxel-based networks suffer from a cubic increase in memory consumption with resolution scaling, while point-based networks are hindered by irregular memory access patterns and expensive nearest neighbor searches. PVCNN mitigates these limitations through a dual-branch architecture, as shown in Fig. 3. The voxel-based branch normalizes the input point cloud by translating it to a local coordinate system centered at its gravity center, then scaling it to fit within a unit sphere. After normalization, the point cloud undergoes voxelization, where points are mapped onto a regular voxel grid by averaging their features within each voxel. This structured representation enables efficient 3D convolutions, allowing the model to aggregate coarse-grained local spatial correlations while maintaining computational efficiency. The point-based branch, on the other hand, retains high-resolution pointwise features by applying a Multi-Layer Perceptron (MLP) to individual points. This direct point-wise transformation ensures the preservation of fine-grained geometric details, which are often lost in voxelized representations due to discretization errors. After convolutional processing, the voxel-based features are devoxelized, where they are mapped back to individual points using trilinear interpolation to ensure distinct feature representations for each point. Finally, the fusion step integrates the outputs of both branches, combining the coarse spatial context from the voxel-based branch with the fine-grained details from the point-based branch. This hybrid design enables PVCNN to preserve fine-grained geometric details while efficiently aggregating neighborhood information, making it well-suited for our model. In our PCD-Liver framework, where the generative model progressively refines point distributions over iterative diffusion steps, PVCNN's ability to efficiently capture both local and global structures is particularly advantageous. By fusing pointwise and voxel-based features, PVCNN ensures accurate and efficient modeling of the complex spatial relationships inherent in point cloud data, making it a powerful backbone for both generative tasks and rigid alignment processes.

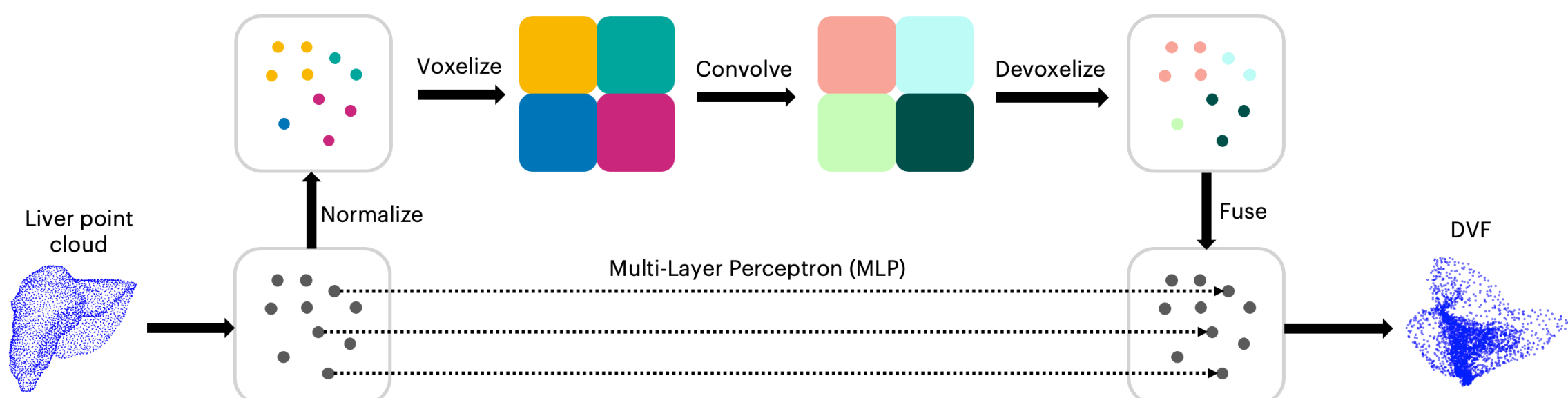

**Figure 3.** Illustration of the PVCNN's two-branch design, wherein the voxel-based branch performs structured 3D convolutions to capture coarse local spatial correlation, while the point-based branch maintains high-resolution pointwise features with minimal memory overhead.

### *2.3.4 Loss function design*

We trained the rigid alignment model by the MSE loss, and trained the point cloud diffusion model using a combined loss function of a denoising loss and a liver surface point cloud similarity loss. The denoising loss $L_{noise}$ is defined as the MSE between the predicted noise and the added noise, as shown in Eq. 2. The similarity loss quantifies the distance between the predicted liver surface point cloud, generated from the predicted DVFs, and the 'ground-truth' liver surface point cloud. This loss is composed of three components: a weighted MSE loss, a Laplacian loss, and a deformation energy loss. The weighted MSE loss $L_{sim}$ calculates the L2 distance between the predicted and 'ground-truth' target liver surface point clouds along the x, y, and z axes. Based on previous reports (Yoganathan *et al.*, 2017), respiration-induced liver motion predominantly occurs along the superior-inferior (SI) axis. Therefore, the MSE loss in the SI direction is scaled by a factor of 3, as concluded from a parameterized study in our prior work (Shao *et al.*, 2023b), to improve the model performance. The Laplacian loss $L_{\mathrm{lap}}$ acts as a regularization term for the liver surface DVFs, preventing unrealistic large local deformations (Wang *et al.*, 2018). It computes the mean square difference between the Laplacian coordinates before ($\boldsymbol{\delta}_p$) and after ($\boldsymbol{\delta}_p^{'}$) deformation, where $\boldsymbol{\delta}_p$ is the Laplacian coordinate of a node $p$ in $\boldsymbol{X}_t$ and $\boldsymbol{\delta}_p^{'}$ is the Laplacian coordinate of the same node in $\boldsymbol{X}_{t-1}$, for $t \in [0, T]$:

$$L_{\text{lap}} = \frac{1}{N}\sum || \boldsymbol{\delta}_p^{'} - \boldsymbol{\delta}_p ||^2, \tag{5}$$

where N denotes the total node number. The Laplacian coordinates, which describes how much the position of the node $p$ deviates from the average position of its neighboring nodes $q$, for each node $p$ are defined by:

$$\boldsymbol{\delta}_p = p - \sum_{q \in \mathcal{N}(p)} \frac{q}{|\mathcal{N}(p)|}, \tag{6}$$

where $\mathcal{N}(p)$ denotes the neighboring nodes of node $p \in \boldsymbol{X}_t$. In our experiments, we select 8 neighboring nodes for each node, following the settings from Pixel2Mesh (Wang *et al.*, 2018). The deformation energy loss $L_{\text{eng}}$ measures the mean deformation energy of the predicted DVFs to regularize the smoothness of the liver surface point cloud deformation:

$$L_{\text{eng}} = \frac{1}{N}\sum k \cdot \left(k - \sum_{q \in \mathcal{N}(k)} \frac{q}{|\mathcal{N}(k)|}\right), \tag{7}$$

where $k \in \boldsymbol{X}_{DVF}^t$ denotes a node in the intermediate liver DVFs. The total loss function is a weighted sum of all the losses:

$$L_{\text{total}} = L_{noise} + \lambda_{sim} L_{sim} + \lambda_{\text{lap}} L_{\text{lap}} + \lambda_{\text{eng}} L_{\text{eng}}, \tag{8}$$

where $\lambda$ defines the weighting coefficients. We used $\lambda_{\text{sim}} = 1$, $\lambda_{\text{lap}} = 0.01$, and $\lambda_{\text{eng}} = 0.01$ after trial-and-error-based optimizations.

#### *2.3.5 Synergy of PCD-Liver with biomechanical modeling for liver tumor localization*

While PCD-Liver effectively tracks the liver surface motion, intra-liver DVFs are needed to accurately localize tumors in low-contrast liver regions. Intra-liver tumors often have low contrast against surrounding liver tissues, making it challenging to extract reliable motion vectors directly from pixel intensities (Zhang *et al.*, 2019). To improve the tracking of low-contrast and/or small anatomical structures, previous works (Shao *et al.*, 2022; Zhang, 2021) have combined deep learning with biomechanical modeling. Such approach estimates deformation fields at high-contrast anatomical boundaries and propagates them into low-contrast regions using finite element analysis (FEA) (Zhang *et al.*, 2019; Zhang *et al.*, 2017), enabling accurate intra-liver tumor localization. In this study, we integrated PCD-Liver with a deep learning-based biomechanical model (Bio model) (Shao *et al.*, 2023b) to eliminate the need for time-consuming iterative optimization, leveraging domain knowledge of tissue biomechanics and finite element analysis for liver tumor localization. Specifically, the liver boundary DVFs derived from PCD-Liver were used as boundary conditions for the deep learning-based liver biomechanical model, enabling the estimation of intra-liver motion.

The Bio model was based on the U-Net architecture (Ronneberger *et al.*, 2015). To enhance feature learning, the liver boundary DVFs obtained from PCD-Liver were decomposed into two components: a spatially uniform (DC) component, representing rigid motion, and a residual oscillating (AC) component, capturing non-rigid deformation. Only the AC component was inputted into the U-Net, while the DC component was later added back to reconstruct the complete intra-liver DVFs. Since convolutional networks require regular Euclidean input data, the AC component, defined on a boundary mesh, was gridded into a volumetric representation via nearest-neighbor interpolation before being processed by the U-Net. The Bio model leveraged a Mooney–Rivlin material model (Shao *et al.*, 2021; Zhang *et al.*, 2019) to generate ground-truth intra-liver DVFs for model training, promoting physically realistic motion estimation. By integrating the Bio model, we effectively translated PCD-Liver-derived surface motion into detailed intra-liver deformation fields, achieving liver tumor localization.

## 2.4 Experimental setup

### 2.4.1 Dataset curation

We followed a similar dataset curation and augmentation procedure as in the published work (Shao *et al.*, 2023b). A dataset of 10 liver cancer patients was used to train and evaluate the proposed patient-specific framework, under an approved IRB protocol. The numbers of nodes of the liver surface point cloud for those patients range from 826 to 5092. Each patient underwent a 10-phase 4D-CT scan, which was resampled to a 2 $mm^3$ isotropic resolution for a matrix size of $256 \times 256 \times 128$. To train the patient-specific model, data augmentations were implemented to simulate realistic, respiration-induced liver motion. This augmentation process employed a PCA-based motion model with the following key steps: (i) performing deformable image registration on 4D-CT using Elastix (Klein *et al.*, 2010) to obtain the volumetric DVFs between the reference phase (0% phase) and the other phases (10%–90%). Registration accuracy was improved by enhancing liver boundaries using a liver-density override technique to improve boundary deformation accuracy. (ii) The liver boundary DVFs were extracted and then fed into a finite-element solver via the FEBio package (Maas *et al.*, 2012) to generate physics-plausible intra-liver DVFs (Shao *et al.*, 2021). (iii) Constructing a patient-specific motion model through PCA on the DVFs, representing respiratory motion as a linear combination of the mean DVF $\boldsymbol{D}_0$ and the first three principal motion components $\boldsymbol{D}_i (i = 1\text{–}3)$, expressed as:

$$\boldsymbol{D}(\boldsymbol{x}, t\%) = \sum_{i=0}^{3} w_i(t\%) \times \boldsymbol{D}_i(\boldsymbol{x}), \tag{9}$$

where $\boldsymbol{x}$ denotes the voxel index, and $w_i(t\%)$ represents the weighting of the mean or principal motion components at respiratory phase $t\%$. (iii) Deformable augmentation was applied by varying the coefficients $w_i(t\%)$ within predefined ranges: $[0.95, 1.05]$ for $w_0$, $[-1.5, 3.0]$ for $w_1$ and $w_2$, and $[-1.5, 1.5]$ for $w_3$. For each respiratory phase $t\%$, 4, 6, 4, and 2 random scaling factors were sampled uniformly from these ranges, producing deformed CT volumes based on the reference-phase CT ($0\%$ phase). The above deformable augmentation strategy was applied to phases $10\%-40\%$ and $70\%-80\%$ of the original 4D-CT scans to generate 768 cases for training and 384 cases for validation, respectively.

To evaluate the model's generalization to unseen motion scenarios, for testing, the phases $50\%-60\%$ and $90\%$ were used due to their larger motion relative to other phases, including the end-of-exhale and end-of-inhale phases. To generate testing motion scenarios more deviating from the training and validation sets to avoid potential data leakage and also to test the model robustness, $w_0$ was set to 1 (compared to $[0.95, 1.05]$ of training/validation), $w_1$ and $w_2$ were sampled from [0.0,4.0] (compared to $[-1.5, 3.0]$ of training/validation), and $w_3$ was sampled from [0.0,2.0] (compared to $[-1.5, 1.5]$ of training/validation) ranges, respectively. For each testing respiratory phase, we sampled 15 combinations of $w_1$, $w_2$, and $w_3$, resulting in 45 testing cases for each patient.

To further enhance robustness, random 3D translations in the range of $[-6.0mm, 6.0mm]$ were applied to LR, AP, and SI directions in training/validation, and $[0.0mm, 10.0mm]$ for testing, to simulate onboard rigid motion and patient setup errors. Each deformation-augmented volume generated three additional volumes from the random translations. The resulting dataset comprised 3072 volumes for training, 1536 for validation, and 135 for testing (with the zero-rigid-shift CT volume discarded during testing, to ensure each testing case involves both deformable motion and rigid translation).

Figure 4(a) illustrates the distribution of liver boundary center-of-mass respiratory motion ranges across ten patients, separated by training, validation, and test subsets. The box plots demonstrate that our dataset captures a broad range of deformable motion patterns, indicative of varying respiratory behaviors and anatomical differences among patients. Notably, the motion range in the test set is intentionally enlarged than those in the training and validation sets, aiming to evaluate the model robustness under unseen motion conditions. The mean liver tumor motion magnitudes (in vector length) across the three subsets are 10.89±4.95 mm (train), 9.57±4.56 mm (validation), and 15.70±6.96 mm (test), and the mean liver boundary motion magnitudes are 7.77±2.52 mm (train), 6.20±2.44 mm (validation), and 11.18±3.56 mm (test), confirming an increasing trend in motion complexity from training to testing (Table I). This distributional shift challenges the model to generalize beyond the training distribution. Additionally, to simulate inter-fractional motion, random rigid translations were applied to the whole volume during data preparation: [–6.0 mm, 6.0 mm] for training/validation and [0.0 mm, 10.0 mm] for testing. The tumor volumes in our dataset range from 0.48 cm³ to 24.62 cm³, with an average volume of 7.74 ± 8.16 cm³, representing a broad distribution of tumor sizes.

(a)

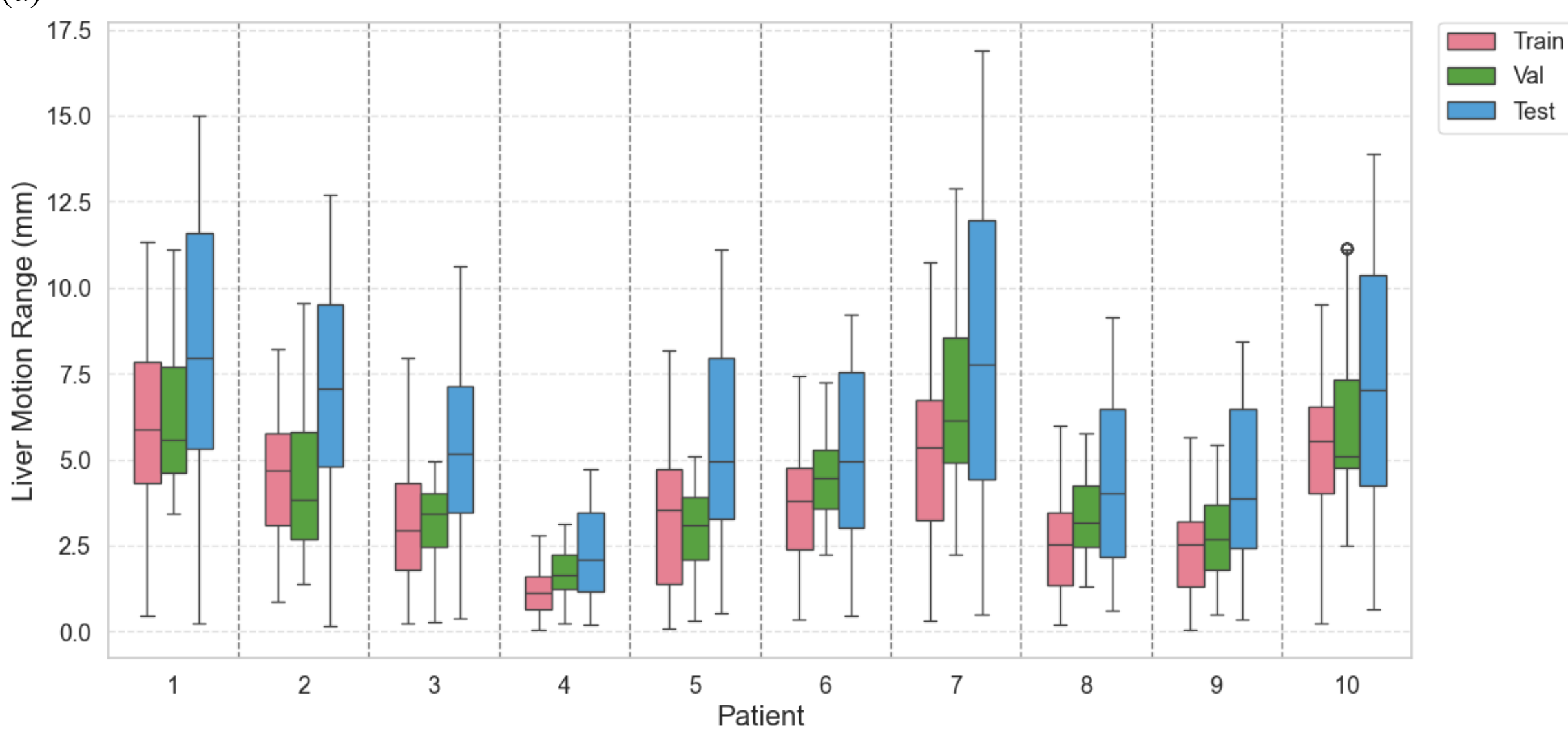

(b)

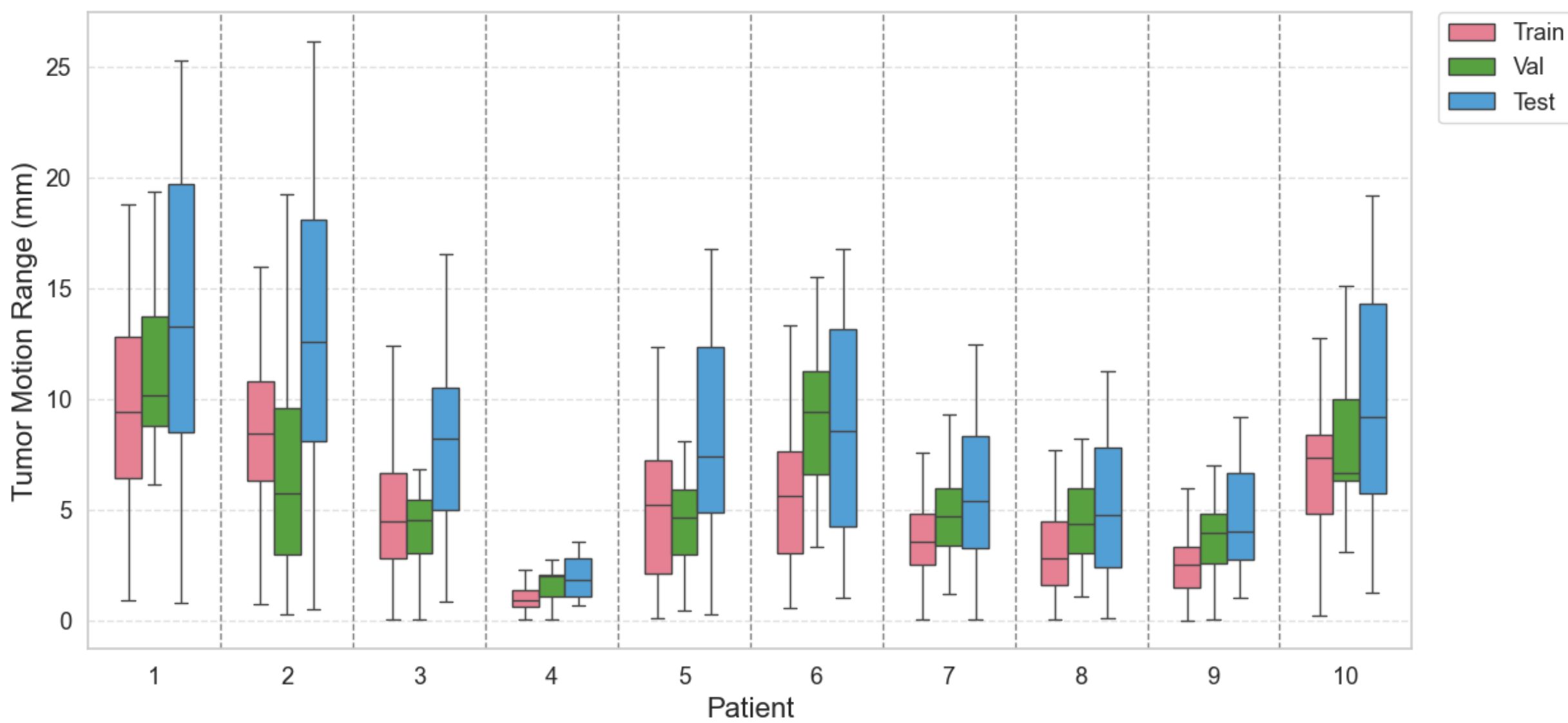

**Figure 4.** (a) Distribution of liver boundary center-of-mass respiratory motion ranges (in mm) for the simulated dataset. (b) Distribution of liver tumor center-of-mass motion ranges (in mm) for the simulated dataset.

**Table I.**

**(a)** Mean ranges of the liver boundary center-of-mass respiratory motion in millimeters (mm).

| Set | SI | | | AP | | | LR | | |
|---|---|---|---|---|---|---|---|---|---|
| | Mean(±std.) | Min | Max | Mean(±std.) | Min | Max | Mean(±std.) | Min | Max |
| Training | 6.83±2.37 | 2.47 | 10.38 | 3.40±1.19 | 1.01 | 5.31 | 1.20±0.68 | 0.27 | 2.22 |
| Validation | 5.57±2.25 | 2.71 | 9.70 | 2.72±1.23 | 0.75 | 4.40 | 0.98±0.40 | 0.52 | 1.84 |
| Test | 9.91±3.27 | 4.24 | 14.55 | 4.78±1.74 | 1.58 | 8.30 | 1.64±0.91 | 0.24 | 2.89 |

**(b)** Mean ranges of the liver tumor center-of-mass respiratory motion in millimeters (mm).

| Set | SI | | | AP | | | LR | | |
|---|---|---|---|---|---|---|---|---|---|
| | Mean(±std.) | Min | Max | Mean(±std.) | Min | Max | Mean(±std.) | Min | Max |
| Training | 9.41±4.75 | 1.87 | 17.28 | 4.63±2.91 | 1.33 | 10.42 | 2.16±1.06 | 0.32 | 3.66 |
| Validation | 8.36±3.73 | 2.62 | 15.73 | 4.39±3.43 | 1.04 | 10.69 | 1.50±1.08 | 0.27 | 3.39 |
| Test | 13.62±6.42 | 2.94 | 23.35 | 6.54±4.59 | 2.10 | 15.95 | 2.70±1.46 | 0.20 | 4.45 |

We calculated the contrast-to-noise ratio (CNR) of the reference phase (0% phase) CT since all other volumes of the same patient in the dataset were derived by deforming this reference image; thus, their intensities remain consistent across train, validation, and test sets. The CNR of the liver relative to the surrounding soft tissues was computed using the following formula:

$$CNR = \frac{\bar{I}_{target} - \bar{I}_{background}}{\sigma_{background}}, \tag{10}$$

where $\bar{I}_{target}$ is the mean intensity of a liver region of interest (ROI), $\bar{I}_{background}$ and $\sigma_{background}$ are the mean and standard deviation of a background ROI located in the nearby soft tissues surrounding the liver. We evaluated the CNR for the inferior portions of the liver by placing a spherical ROI (radius = 5 mm) there. Another spherical ROI was placed in a homogeneous soft tissue region near the inferior liver surface to serve as a background reference, by explicitly avoiding air or organ boundaries. An example of ROI placement is illustrated in Fig. 5. The resulting average CNR for the liver ROIs was 1.94 ± 1.44 across the whole dataset.

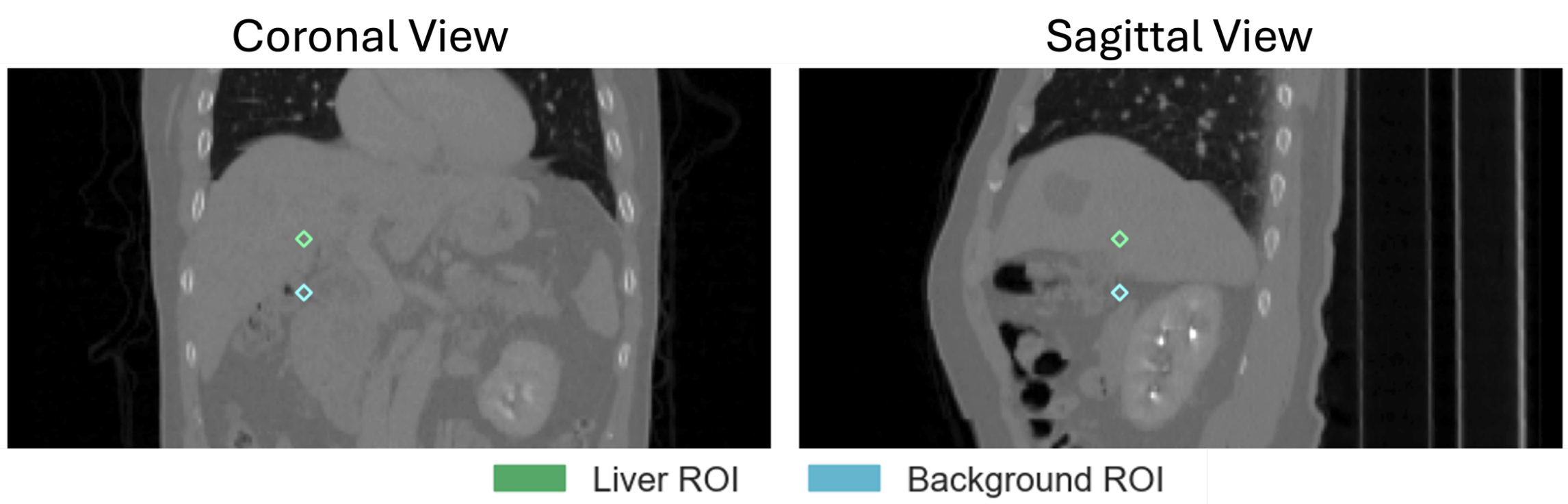

**Figure 5.** Visualization of the regions of interest (ROIs) in coronal and sagittal views used to compute the contrast-to-noise ratio (CNR) of the liver.

*2.4.2 Training setup*

For each of the CT volume, X-ray projections were simulated on-the-fly using the PYRO-NN package (Syben *et al.*, 2019), randomizing projection angles uniformly between 0° and 360° to train/validate the model to generalize across angles. Each projection has $512 \times 384$ pixels with a pixel size of $0.776 \times 0.776\ mm^2$. During training, X-ray quantum noise and electronic noise were added to the simulated projections to enhance model robustness. Photon quantum noise was modeled using a Poisson distribution with a mean of $10^5$ photons per X-ray detector pixel, while electronic noise followed a Gaussian distribution with a standard deviation of 10 photons (Zhang *et al.*, 2013).

Our training approach consists of two stages. First, we train the rigid alignment model (Sec. 2.3.2), using the prior reference liver surface point cloud and the X-ray projections as input. The 'ground-truth' rigid motion, derived from the training DVFs (Sec. 2.4.1), was used for supervision. The ResNet-50 within the geometry-informed feature pooling layer for the rigid alignment model was simultaneously trained. The rigid alignment model training stage runs for 100 epochs. In the second stage, we freeze the rigid alignment model and train a conditional point cloud diffusion model, integrating the shift predicted by the rigid alignment model at each denoising step. For the point cloud diffusion model, the 'ground-truth' training DVFs and the resulting target liver surface point clouds are used for supervision. The diffusion model is trained for 200,000 steps. The model implementation was done in PyTorch (Paszke *et al.*, 2019), and all training and testing were conducted on an Nvidia A100 GPU.

*2.4.3 Evaluation schemes and ablation studies*

We evaluate the liver surface motion estimation accuracy with two metrics: the root mean square error (RMSE) and 95-percentile Hausdorff distance (HD95). Specifically, RMSE calculates the average node-wise distance between the predicted and 'ground-truth' target liver surface point cloud as follows:

$$\mathrm{RMSE} = \sqrt{\frac{1}{N}\sum_{p=1}^{N} || \boldsymbol{X}_{pred}^{p} - \boldsymbol{X}_{tar}^{p} ||^2}, \quad (11)$$

where $N$ is the number of liver surface nodes, and $\boldsymbol{X}_{pred}^{p}$ and $\boldsymbol{X}_{tar}^{p}$ represent the coordinates of the $p$th node on the predicted and 'ground-truth' target liver surface point clouds, respectively. HD95, on the other hand, quantifies the maximum distance between the predicted liver surface point cloud $\boldsymbol{X}_{pred}$ and the 'ground-truth' surface point cloud $\boldsymbol{X}_{tar}$, defined as:

$$\mathrm{HD95} = \max\{P_{95}d(\boldsymbol{x}_{pred}, \boldsymbol{X}_{tar}), P_{95}d(\boldsymbol{x}_{tar}, \boldsymbol{X}_{pred})\}, \quad (12)$$

where $P_{95}$ is the 95th-percentile operator to account for outliers, $\boldsymbol{x}_{pred} \in \boldsymbol{X}_{pred}$ is a node in the predicted liver point cloud, $\boldsymbol{x}_{tar} \in \boldsymbol{X}_{tar}$ is a node in the 'ground-truth' target liver point cloud, and the surface distance between a node $\boldsymbol{x}_{pred} \in \boldsymbol{X}_{pred}$ and the 'ground-truth' target liver point cloud $\boldsymbol{X}_{tar}$ is defined as:

$$d(\boldsymbol{x}_{pred}, \boldsymbol{X}_{tar}) = \min_{x_{tar} \in \boldsymbol{X}_{tar}} || \boldsymbol{x}_{pred} - \boldsymbol{x}_{tar} ||, \quad (13)$$

which measures the closest distance from a node in $\boldsymbol{X}_{pred}$ to any node on the $\boldsymbol{X}_{tar}$. Similarly, $d(\boldsymbol{x}_{tar}, \boldsymbol{X}_{pred})$ measures the closest distance from a node in $\boldsymbol{X}_{tar}$ to the predicted surface point cloud $\boldsymbol{X}_{pred}$, defined as:

$$d(\boldsymbol{x}_{tar}, \boldsymbol{X}_{pred}) = \min_{x_{pred} \in \boldsymbol{X}_{pred}} || \boldsymbol{x}_{tar} - \boldsymbol{x}_{pred} ||, \quad (14)$$

To further examine the potential of the developed approach in tumor localization, we converted the surface cloud of the liver into a volumetric mesh and utilized a deep learning-based biomechanical model (Bio) (Shao *et al.*, 2023b) to infer the inner tumor motion (Sec. 2.3.5). Accuracy of the intra-liver motion estimation was evaluated by the liver tumor localization error, quantified using the center-of-mass error (COME). The COME measures the center-of-mass distance between the predicted and 'ground-truth' liver tumor contours.

For comparison, we also trained the X360 model (Shao *et al.*, 2023a) using the same dataset. To further evaluate the robustness of the diffusion model-based method, we conducted a robustness study under varying noise levels and CNR conditions. For noise robustness, additional photon noise was introduced during the inference stage of PCD-Liver and X360. This noise was modeled using a Poisson distribution with mean values of $10^3$, $10^4$, $10^5$, and $10^6$ photons per X-ray detector pixel. For contrast robustness, we tested PCD-Liver and X360 on test datasets (without changing the CNRs for the training and testing sets) with simulated CNRs of 0.36 ± 0.19, 0.59 ± 0.14, 0.97 ± 0.72, 1.55 ± 1.15, and 2.33 ± 1.73 by adjusting the intensity difference between the liver and background regions.

To assess the effectiveness and contributions of the rigid alignment model and the geometry-informed feature pooling layer in the PCD-Liver framework, we conducted ablation studies by systematically removing these components. First, we developed a variant, PCD-Liver-NR, in which the rigid alignment model was excluded, relying solely on the diffusion model to estimate liver rigid and deformable motion. In this setting, the intermediate DVFs predicted by the diffusion model were not updated by rigid shift estimated from the rigid alignment model. Second, we introduced PCD-Liver-NG, another variant without the geometry-informed feature pooling layer, using only ResNet-50 to extract features from X-ray scans for conditioning the diffusion model. Here, the ResNet-50 output was directly concatenated with the prior reference liver point cloud coordinates to train the rigid alignment model, and with the intermediate liver surface point cloud coordinates when training the diffusion models, respectively. Both PCD-Liver-NR and PCD-Liver-NG were trained using the same dataset and hyper-parameters as the original PCD-Liver model for a fair comparison. For testing, each patient's 135 CT volumes in the test set were used, and nine X-ray projections at equally spaced gantry angles across 360° were simulated, resulting in 1215 testing projections per patient.

We evaluated the statistical significance of registration performance differences between X360 as well as PCD-Liver ablation models to PCD-Liver using the Wilcoxon matched-pairs signed-rank test. This test was chosen because the underlying error distributions are unknown and may not follow a normal distribution.

Additionally, the registration/localization errors were paired across models. Thus, the nonparametric Wilcoxon matched-pairs signed-rank test was employed (Corder and Foreman, 2011).

## 3. Results

### *3.1 Accuracy of the predicted liver surface motion*

This section compares the accuracy of liver surface motion estimation by PCD-Liver and X360. Fig. 6 shows the projection of the prior reference liver point cloud and the PCD-Liver deformed liver surface onto the X-ray projections at five different angles (-120°, -80°, 0°, 80°, and 120°) for two patients (P1 and P2). In each case, the first column displays the prior reference liver surface point cloud (green), while the second column shows the PCD-Liver-deformed liver surface point cloud (red). The arrows point to the regions that show the prior reference liver surface clouds were deformed by the PCD-Liver model to closely align with the underlying anatomy in the X-ray projections. The deformed liver clouds exhibit substantially improved alignment with the liver surfaces seen across different projection angles, demonstrating the model's capability in accurately capturing liver's deformable motion.

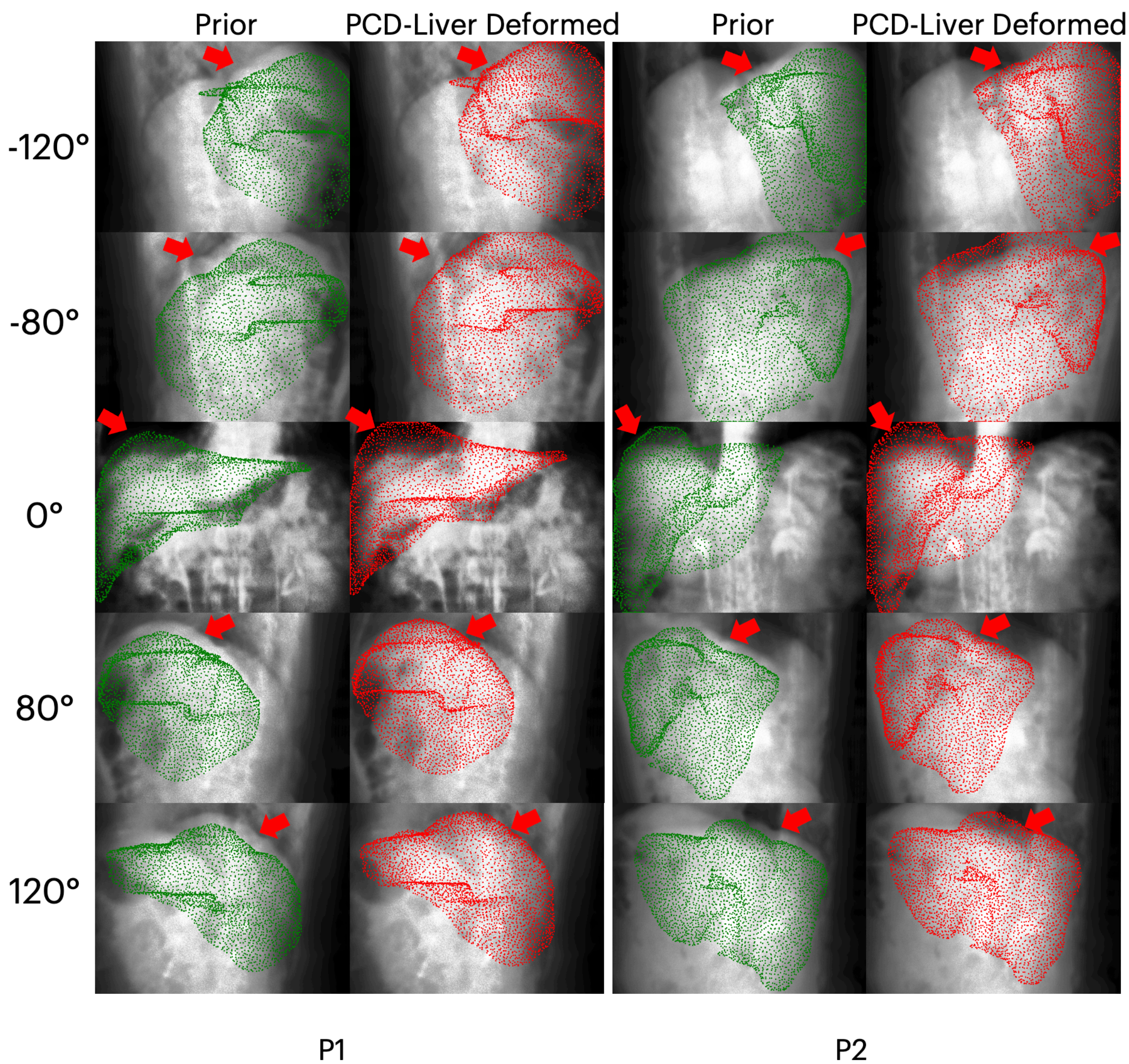

**Figure 6.** Projections of liver surface point clouds onto onboard X-ray projections at -120°, -80°, 0°, 80°, and 120° angles for two patients. The first and the second columns of each patient case present the projections of the liver surface point clouds before and after the PCD-Liver motion inference, respectively. Red arrows point to noticeable deformation regions.

Figure 7 shows a comparison of the liver tumor surface meshes obtained with the deep learning-based Bio model (Sec. 2.3.5), based on the liver boundary motion solved by PCD-Liver and X360 for two patient cases (P1 and P2), respectively. The first column displays the overlays of the prior tumor mesh (green), the target 'ground-truth' tumor mesh (blue), and the PCD-Liver-deformed tumor mesh (red). The second column presents the corresponding overlays for X360. The PCD-Liver model achieves better tumor localization accuracy, by aligning the predicted tumor mesh more closely with the target 'ground- truth' compared to X360.

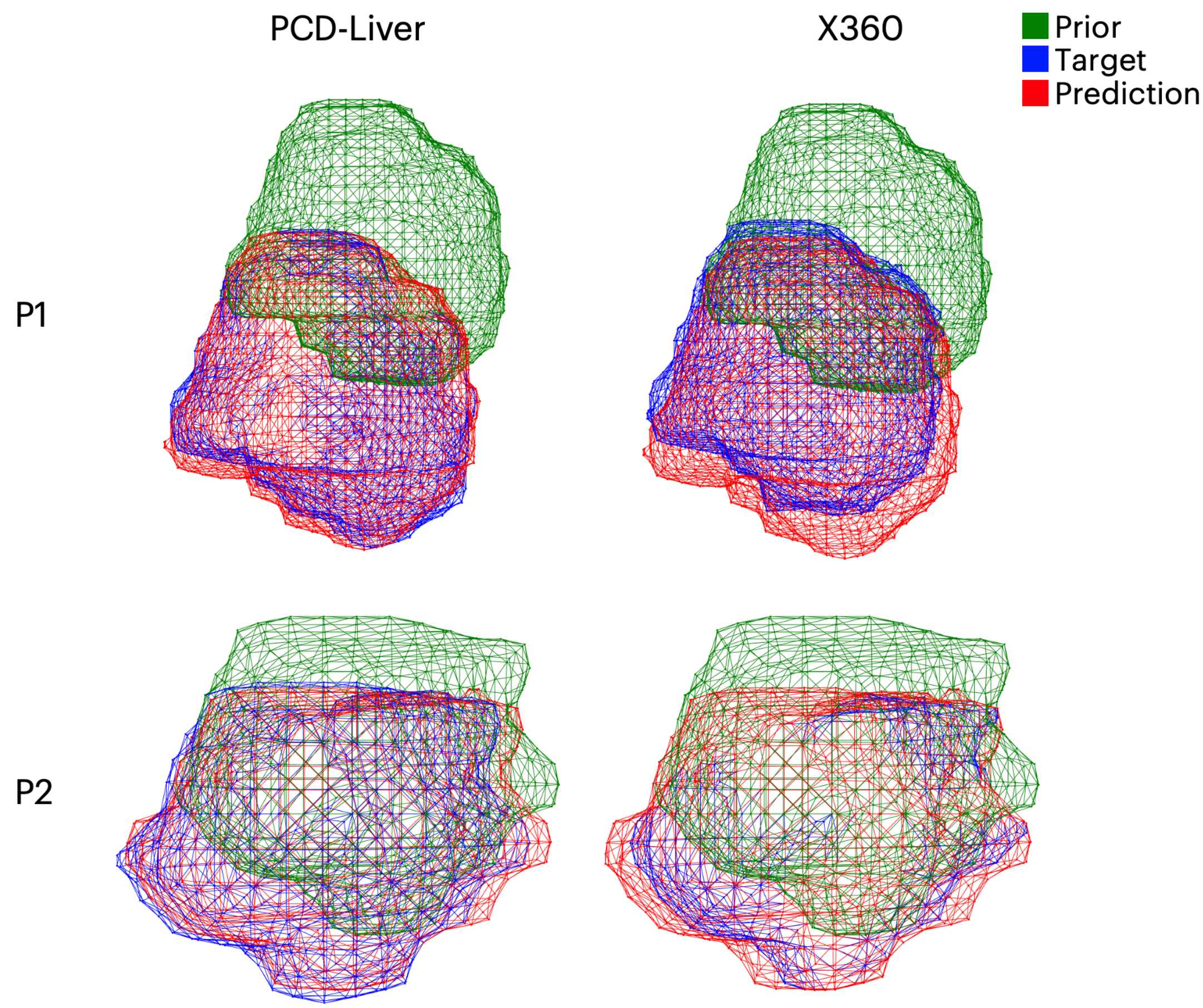

**Figure 7.** Liver tumor overlays between the prior, target, and predicted tumor meshes by PCD-Liver and X360 for two patients.

Table II presents a quantitative comparison between X360 and PCD-Liver in terms of RMSE (liver), HD95 (liver), and COME (tumor) across ten patient cases. The results demonstrate that PCD-Liver consistently outperforms X360, achieving better metric values with smaller standard deviations, indicating improved accuracy and robustness. Specifically, the RMSE and HD95 values between the predicted and target 'ground-truth' liver surface point clouds are consistently lower for PCD-Liver across all patients.

Additionally, COME values, which assess the tumor center-of-mass error, show that PCD-Liver provides more accurate tumor localization.

**TABLE II.** Average (±s.d.) RMSE (liver), HD95 (liver), and COME (tumor) results on the test set by different methods. The arrows are pointing in the direction of improved accuracy. [Prior/Pred, Target] means a corresponding metric calculated between the prior/predicted and the 'ground-truth' target liver meshes. * Indicates statistical tests with $p < 0.05$.

| ID | Model | RMSE (mm) ↓ [Prior, Target] | RMSE (mm) ↓ [Pred, Target] | HD95 ↓ [Prior, Target] | HD95 ↓ [Pred, Target] | COME (mm) ↓ [Prior, Target] | COME (mm) ↓ [Pred, Target] |
|---|---|---|---|---|---|---|---|
| 1 | X360 | 10.80±3.94 | 4.47±1.82 | 14.67±5.32 | 5.00±1.58 | 14.48±6.10 | 4.05±2.17 |
| | PCD-Liver | | **3.54±1.48** | | **4.25±1.14** | | **3.76±1.87** |
| 2 | X360 | 9.73±3.47 | 4.92±2.32 | 12.00±4.19 | 5.77±2.73 | 13.65±5.81 | 6.14±3.61 |
| | PCD-Liver | | **3.70±2.33** | | **4.44±2.63** | | **5.81±3.42** |
| 3 | X360 | 7.83±2.90 | 4.21±1.63 | 8.72±2.97 | 4.66±1.41 | 9.53±3.77 | 4.19±2.15 |
| | PCD-Liver | | **3.29±2.22** | | **3.61±1.68** | | **3.51±2.05** |
| 4 | X360 | 6.71±2.37 | 5.04±1.93 | 7.14±2.17 | 5.42±1.61 | 4.85±2.52 | 3.19±1.81 |
| | PCD-Liver | | **3.04±1.81** | | **3.54±1.61** | | **2.75±1.98** |
| 5 | X360 | 8.46±3.16 | 4.75±1.89 | 9.91±3.60 | 5.41±1.67 | 9.64±3.99 | 4.47±2.11 |
| | PCD-Liver | | **4.22±2.03** | | **4.51±1.15** | | **3.80±1.53** |
| 6 | X360 | 8.61±2.77 | 4.57±1.58 | 11.12±3.41 | 5.78±1.75 | 9.67±4.32 | 4.46±2.33 |
| | PCD-Liver | | **3.71±1.44** | | **4.85±1.85** | | **3.65±1.51** |
| 7 | X360 | 11.35±4.27 | 5.06±2.32 | 14.83±5.40 | 5.78±2.29 | 7.49±3.21 | 3.14±1.79 |
| | PCD-Liver | | **3.67±1.71** | | **3.29±1.55** | | **2.65±1.77** |
| 8 | X360 | 8.03±2.86 | 4.44±1.87 | 10.51±3.34 | 6.15±2.06 | 7.15±3.18 | 3.36±1.89 |
| | PCD-Liver | | **3.98±1.22** | | **5.23±1.42** | | **2.96±1.51** |
| 9 | X360 | 7.35±2.65 | 4.37±1.79 | 8.20±2.58 | 5.06±1.62 | 6.44±2.73 | 3.00±1.72 |
| | PCD-Liver | | **3.48±1.84** | | **4.45±1.47** | | **2.35±1.65** |
| 10 | X360 | 9.75±3.52 | 4.43±1.62 | 11.79±3.57 | 5.87±1.52 | 11.23±4.41 | 3.86±1.96 |
| | PCD-Liver | | **3.55±1.42** | | **4.75±1.41** | | **3.32±1.35** |
| Avg | X360 | 8.82±3.58 | 4.63±1.91* | 10.84±4.55 | 5.48±2.11* | 9.72±4.34 | 3.99±2.39* |
| | PCD-Liver | | **3.63±1.88** | | **4.29±1.75** | | **3.46±2.15** |

### 3.2 Robustness Study

We tested the robustness of X360 and PCD-Liver under different noise levels in X-ray projections. Figure 8 shows that PCD-Liver consistently demonstrates lower RMSE values compared to X360 across all noise levels. At the highest noise level ($10^3$ photons per pixel), X360 exhibits a substantial increase in RMSE, suggesting a notable decline in performance under high noise. Conversely, PCD-Liver maintains a relatively stable RMSE, indicating better robustness to noise.

Figure 9 shows the projection of the target 'ground-truth' liver surface point cloud and the predicted liver surface point cloud from X360 and PCD-Liver onto the X-ray images under different noise levels. Even as noise levels increase from $10^6$ to $10^3$ photons per pixel, PCD-Liver produces consistently stable predictions of liver boundary movement, even for the highest noise level ($10^3$) while the liver boundary is corrupted by the excessive noise unseen in the training set. In contrast, X360 shows substantial susceptibility to increased noise levels, especially to the $10^3$ and $10^4$ noise levels which deviate from the training dataset where noises based on a $10^5$ photons/pixel dose level were simulated (Sec. 2.4.2). At $10^3$ and $10^4$ noise levels, the yellow dashed reference line indicates that X360 underperforms in rigid alignment, failing to accurately match the liver's position to the target mesh, whereas PCD-Liver maintains alignment despite the high noise. Additionally, the red arrows highlight key regions where X360 fails to deform the prior liver structure to account for deformable motion, leading to visible mismatches between the predicted and actual liver meshes. In contrast, PCD-Liver effectively captures the deformable motion, ensuring that the predicted liver surface closely follows the ground-truth liver mesh, even under severe noise conditions. This suggests that PCD-Liver is more robust to both rigid and non-rigid deformations, maintaining structural integrity and anatomical consistency across varying noise levels.

Table III shows the average (±s.d.) RMSE scores of the deformed liver surface meshes generated by the X360 and PCD-Liver models across a range of liver CNRs. The CNRs were only varied for the testing dataset (no change to the training and validation sets), to evaluate the models' robustness to data distribution variations. The original training dataset had an average CNR of 1.94 ± 1.44. To systematically assess the models' robustness to contrast variations, we simulated testing datasets with varying CNRs: 0.36 ± 0.19, 0.59 ± 0.14, 0.97 ± 0.72, 1.55 ± 1.15, and 2.33 ± 1.73, by changing the intensity differences between the liver and the surrounding tissues in the reference phase volume, before applying augmentations to generate the test datasets. While X360 maintained moderate accuracy at higher mean CNR levels, it shows a substantial performance degradation in low-contrast settings, with RMSE rising to 10.29 ± 4.24 mm at an average CNR of 0.36. In contrast, PCD-Liver demonstrated consistently low RMSE values across all contrast conditions, with minimal performance variance (ranging from 3.62 to 4.01 mm), highlighting its superior robustness to contrast variability. This robustness of PCD-Liver against noise and contrast variations is likely due to the diffusion model's inherent resilience to out-of-distribution (OOD) data.

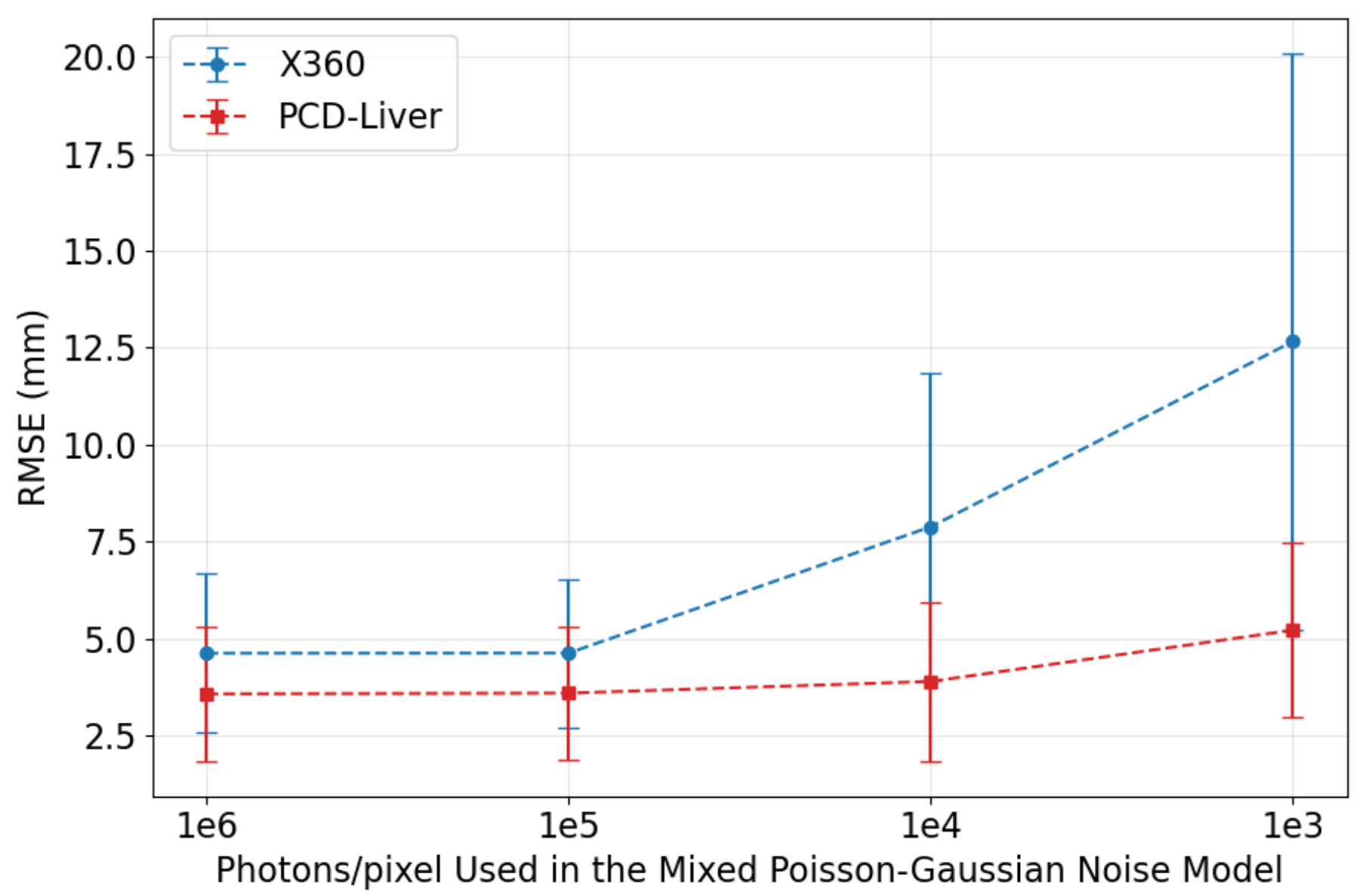

**Figure 8.** RMSE comparison between X360 and PCD-Liver under varying noise levels in X-ray projections. The PCD-Liver model consistently demonstrates lower RMSE across all noise levels, indicating superior robustness to noise compared to X360. As noise levels increase, X360 shows a substantial rise in RMSE, while the PCD-Liver maintains relatively stable performance.

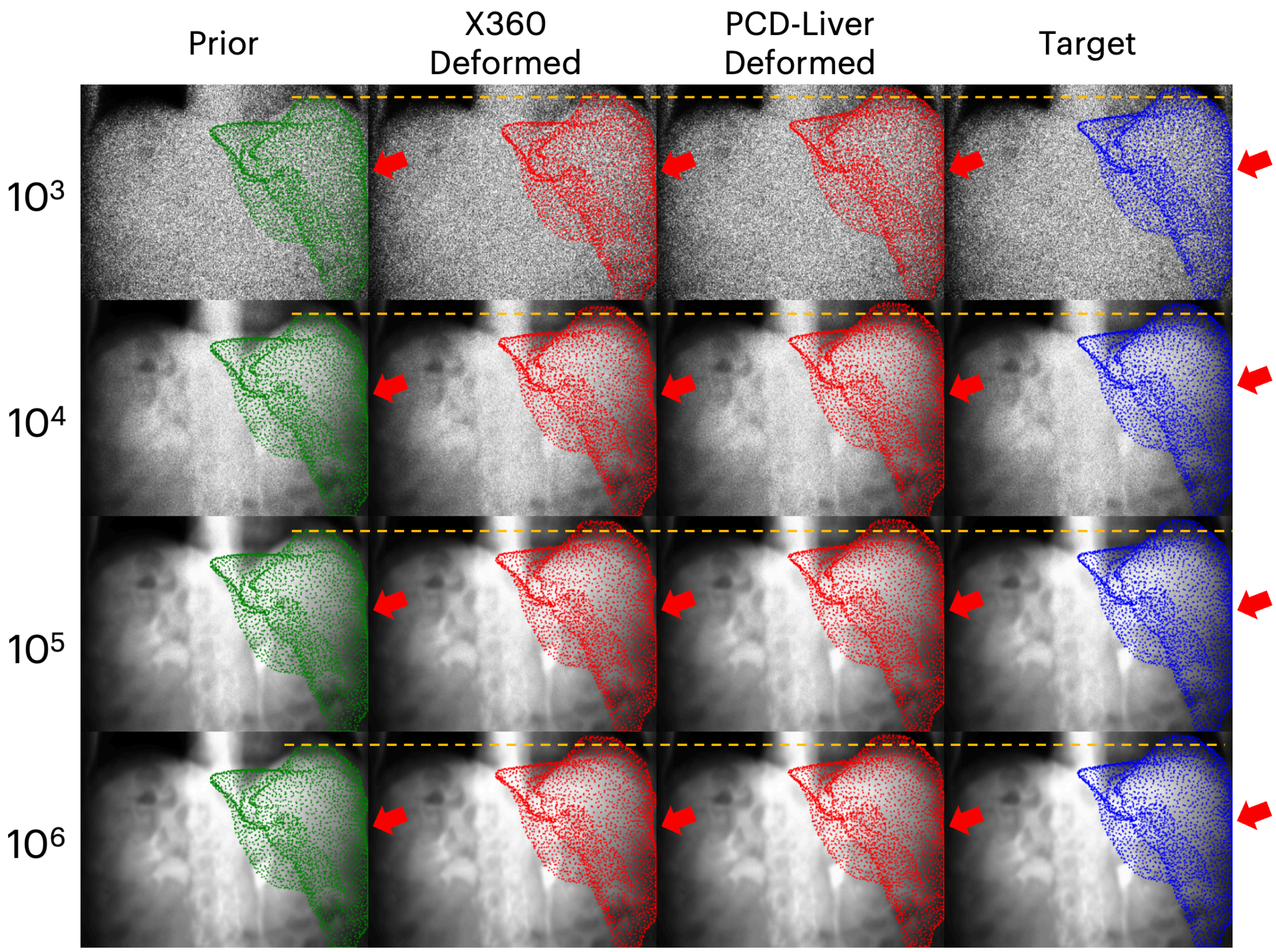

**Figure 9.** Predicted liver surface point clouds of PCD-Liver projected onto the X-ray images of different noise levels ($10^3$, $10^4$, $10^5$, and $10^6$, from top to bottom, respectively). The X-ray images are acquired with a gantry angle of 200°.

**TABLE III.** Average (±s.d.) RMSE (liver) results of PCD-Liver and X360 on the test set with varying mean CNRs.

| Model | RMSE (mm) ↓ [Prior, Target] | RMSE (mm) ↓ [Pred, Target] | | | | | |
|---|---|---|---|---|---|---|---|
| | | CNR: 0.36 | CNR: 0.59 | CNR: 0.97 | CNR: 1.55 | CNR: 1.94 | CNR: 2.33 |
| X360 | 8.82±3.58 | 10.29±4.24 | 6.09±2.88 | 5.10±2.01 | 4.93±1.98 | 4.63±1.91 | 4.58±2.12 |
| PCD-Liver | | **4.01±2.19** | **3.89±2.09** | **3.84±2.12** | **3.73±1.88** | **3.63±1.88** | **3.62±1.72** |

### *3.3 Ablation Study*

Table IV compares PCD-Liver against its two ablated variants: PCD-Liver-NG (without the geometry-informed feature pooling layer) and PCD-Liver-NR (without the rigid alignment model). All $p$-values of the Wilcoxon signed-rank tests between the PCD-Liver, PCD-Liver-NG, and PCD-Liver-NR models reached the significance level ($p < 0.05$). Across all patients and evaluation metrics (RMSE, HD95, and COME), PCD-Liver consistently outperforms its ablated counterparts, demonstrating the importance of both the rigid alignment model and the geometry-informed feature pooling layer. For PCD-Liver-NG, use of non-spatial-encoded ResNet-50 feature vectors instead of the geometry-informed pooling layer leads to an increase in prediction error, indicating that the specialized pooling mechanism provides more relevant features matching to the projection geometry that improve the model accuracy. On the other hand, for PCD-Liver-NR, the removal of the rigid alignment model shows the limitation of the diffusion model in capturing both rigid and deformable motion.

**TABLE IV.** Average (±s.d.) RMSE (liver), HD95 (liver), and COME (tumor) results on the test set by different PCD-Liver variants. The arrows are pointing in the direction of improved accuracy. [Prior/Pred, Target] means a corresponding metric calculated between the prior/predicted and the 'ground-truth' target liver meshes. * Indicates statistical tests with $p < 0.05$.

| ID | Model | RMSE (mm) ↓ [Prior, Target] | RMSE (mm) ↓ [Pred, Target] | HD95 ↓ [Prior, Target] | HD95 ↓ [Pred, Target] | COME (mm) ↓ [Prior, Target] | COME (mm) ↓ [Pred, Target] |
|---|---|---|---|---|---|---|---|
| 1 | PCD-Liver-NG | 10.80±3.94 | 3.99±1.95 | 14.67±5.32 | 4.58±2.42 | 14.48±6.10 | 3.97±1.83 |
| | PCD-Liver-NR | | 4.73±2.31 | | 5.13±2.27 | | 4.34±2.64 |
| | PCD-Liver | | **3.54±1.48** | | **4.25±1.14** | | **3.76±1.87** |
| 2 | PCD-Liver-NG | 9.73±3.47 | 4.36±2.37 | 12.00±4.19 | 5.32±2.78 | 13.65±5.81 | 5.92±3.81 |
| | PCD-Liver-NR | | 4.94±2.26 | | 5.39±2.31 | | 6.16±3.57 |
| | PCD-Liver | | **3.70±2.33** | | **4.44±2.63** | | **5.81±3.42** |
| 3 | PCD-Liver-NG | 7.83±2.90 | 4.13±2.21 | 8.72±2.97 | 4.62±1.84 | 9.53±3.77 | 4.36±1.87 |
| | PCD-Liver-NR | | 4.84±2.23 | | 5.05±2.05 | | 4.86±2.56 |
| | PCD-Liver | | **3.29±2.22** | | **3.61±1.68** | | **3.51±2.05** |
| 4 | PCD-Liver-NG | 6.71±2.37 | 3.68±1.98 | 7.14±2.17 | 4.47±1.75 | 4.85±2.52 | 2.82±2.24 |
| | PCD-Liver-NR | | 4.72±2.07 | | 5.02±1.82 | | 2.95±2.23 |
| | PCD-Liver | | **3.04±1.81** | | **3.54±1.61** | | **2.75±1.98** |
| 5 | PCD-Liver-NG | 8.46±3.16 | 4.12±1.79 | 9.91±3.60 | 4.93±1.50 | 9.64±3.99 | 3.96±1.36 |
| | PCD-Liver-NR | | 4.22±2.03 | | 5.18±1.99 | | 4.09±2.66 |
| | PCD-Liver | | **4.22±2.03** | | **4.51±1.15** | | **3.80±1.53** |
| 6 | PCD-Liver-NG | 8.61±2.77 | 4.16±1.63 | 11.12±3.41 | 4.99±1.62 | 9.67±4.32 | 3.91±2.06 |
| | PCD-Liver-NR | | 4.43±1.94 | | 4.98±1.68 | | 4.26±2.19 |
| | PCD-Liver | | **3.71±1.44** | | **4.85±1.85** | | **3.65±1.51** |
| 7 | PCD-Liver-NG | 11.35±4.27 | 4.25±2.14 | 14.83±5.40 | 4.58±1.60 | 7.49±3.21 | 3.33±1.73 |
| | PCD-Liver-NR | | 5.45±3.07 | | 5.63±2.64 | | 3.77±1.97 |
| | PCD-Liver | | **3.67±1.71** | | **3.29±1.55** | | **2.65±1.77** |
| 8 | PCD-Liver-NG | 8.03±2.86 | 4.38±1.84 | 10.51±3.34 | 5.57±1.90 | 7.15±3.18 | 3.22±1.98 |
| | PCD-Liver-NR | | 4.81±2.12 | | 5.79±2.14 | | 3.51±2.32 |
| | PCD-Liver | | **3.98±1.22** | | **5.23±1.42** | | **2.96±1.51** |
| 9 | PCD-Liver-NG | 7.35±2.65 | 3.97±2.12 | 8.20±2.58 | 4.97±1.93 | 6.44±2.73 | 2.79±2.17 |
| | PCD-Liver-NR | | 4.97±2.20 | | 5.52±1.95 | | 3.42±1.86 |
| | PCD-Liver | | **3.48±1.84** | | **4.45±1.47** | | **2.35±1.65** |
| 10 | PCD-Liver-NG | 9.75±3.52 | 4.14±1.69 | 11.79±3.57 | 5.40±1.82 | 11.23±4.41 | 3.76±1.91 |
| | PCD-Liver-NR | | 4.61±1.93 | | 5.91±1.75 | | 4.08±1.99 |
| | PCD-Liver | | **3.55±1.42** | | **4.75±1.41** | | **3.32±1.35** |
| Avg | PCD-Liver-NG | 8.82±3.58 | 4.17±1.99* | 10.84±4.55 | 4.95±1.89* | 9.72±4.34 | 3.80±2.35* |
| | PCD-Liver-NR | | 4.77±2.26* | | 5.36±2.19* | | 4.14±2.59* |
| | PCD-Liver | | **3.63±1.88** | | **4.29±1.75** | | **3.46±2.15** |

## 4. Discussion

In this work, we proposed the PCD-Liver framework that combines a conditional point cloud diffusion model with a rigid alignment model to improve liver motion estimation using single X-ray projections. Our approach incorporates a geometry-informed feature pooling layer to enhance the relevance of extracted features for motion estimation, while the rigid alignment model ensures spatial consistency in the predicted DVFs by the diffusion model. By integrating these components, the framework achieves robust and accurate liver motion predictions, even under challenging imaging conditions with high noise levels. Although previous works of optical-flow approaches (Teo *et al.*, 2013; Fontaine *et al.*, 2017) share the same goal as PCD-Liver of offering real-time, marker-less tumor tracking, their methodologies diverge substantially. The optical flow pipeline extracts 2D motion vectors from pairs of consecutive projections and integrates these vectors over time to reconstruct the tumor's 2D in-plane trajectory. In contrast, PCD-Liver infers a 3D DVF of the liver surface relative to a reference liver mesh from a single projection, addressing volumetric motion rather than just 2D in-plane shifts, making the task of PCD-Liver inherently more challenging.

PCD-Liver repurposes several existing components into a unified framework to address key challenges of liver tumor motion tracking. While the current state-of-the-art approach (X360) suffers from limited accuracy and a high sensitivity to imaging noise, PCD-Liver maintains high fidelity across a wide range of noise and contrast conditions, making it better suited for real clinical use. Importantly, we diverge from the typical use of diffusion models for unconditional/conditional point cloud generation by reframing the problem as a shape/deformation prediction problem under anatomical constraints. Conditioning the diffusion process on spatial features extracted from single X-ray projections, we predict DVFs that register a prior reference liver surface mesh to its target deformed states. In the setting of conventional shape generation tasks, the primary focus is on preserving overall shape integrity, while the fidelity of global/local geometry is less critical. However, in our setting, even small deviations in predicted DVFs can lead to anatomically implausible mesh deformations, requiring tighter conditioning of the model than conventional shape-generation tasks. Additionally, we identified a limitation of point cloud diffusion models in producing near-zero-mean outputs when tasked with simultaneous deformation and rigid translation. To overcome this, we incorporated a rigid-alignment module that explicitly corrects global translation at every diffusion step, preserving both shape fidelity and real-world spatial positioning, which is essential for downstream tasks like intra-treatment tumor localization and treatment adaptation. Overall, our contributions represent a novel and clinically meaningful adaptation of applying diffusion models for anatomically guided shape/deformation prediction and offering an innovative adoption of generative techniques in high-precision, image-guided therapies.

The model's performance was validated using several evaluation metrics, including RMSE, HD95, and COME, demonstrating that our model consistently outperforms the previous state-of-the-art method, the X360 model, in both accuracy and robustness. As shown in Table II, PCD-Liver achieves an average RMSE improvement of approximately 1 mm and a reduction in COME by 0.5 mm across 10 patient cases. These results confirm that the diffusion-based approach offers more precise liver motion estimation compared to the previous GNN-based X360 model. Furthermore, Figure 8 illustrates the robustness of PCD-Liver under varying noise levels. While X360 exhibits a sharp increase in RMSE at higher noise levels ($10^3$ and $10^4$ photons per pixel), PCD-Liver maintains relatively stable performance, indicating its superior ability to generalize under real-world imaging uncertainties. Figure 9 further supports this finding by showing that, under highly noisy conditions, X360 struggles with rigid alignment and fails to deform the prior liver structure properly, whereas PCD-Liver consistently preserves the liver boundary integrity, effectively handling both rigid motion and non-rigid deformation. Also, as shown in Table III, the performance of PCD-Liver is consistent across different contrast scenarios, while X360 shows significant performance degradation facing low contrast cases, which further suggesting the robustness of the PCD-Liver.

We conducted an ablation study to evaluate the contributions of key components in the PCD-Liver framework. Specifically, we assessed two variants: (1) PCD-Liver-NG, which removes the geometry-informed pooling layer, and (2) PCD-Liver-NR, which removes the rigid alignment model. Overall, the performance hierarchy was: PCD-Liver > PCD-Liver-NG > PC qD-Liver-NR, highlighting the importance of both components. The rigid alignment model was introduced to address a known limitation of the point cloud diffusion model-that it tends to generate near-zero-mean outputs. For deformable registration, it means that the inferred DVFs may not be able to resolve rigid translation correctly. The results from PCD-Liver-NR confirm that without the rigid alignment module, the diffusion model fails to capture both rigid and deformable motion simultaneously, resulting in a significant performance drop. The results indicate that explicit modeling of global translation is essential for accurate liver deformation and shape prediction. On the other hand, the geometry-informed feature pooling layer was designed to provide projection angle-aware conditioning by mapping 2D image features to 3D liver surface nodes using cone-beam geometry. Removing this component (PCD-Liver-NG) and conditioning the diffusion model directly with ResNet-50 features led to degraded performance. This suggests that while the ResNet-50 can learn image features relevant to liver deformation, it lacks geometric awareness and requires the model to implicitly infer projection angle information, which reduces efficiency and precision. These results reinforce the value of direct, physics-based conditioning to better extract relevant features to constrain diffusion models for tasks that demand high spatial accuracy. The performance degradation of PCD-Liver-NG is less pronounced as compared to PCD-Liver-NR (where the rigid module is removed), implying that even without geometric pooling, the diffusion model can somewhat use the extracted image features in a more implicit way. With the implicit learning capability, the model may benefit from additional features from other imaging modalities that lack direct physics-driven correspondence. For instance, incorporating additional structural cues from surface imaging (Shao *et al.*, 2023b), may offer a promising path to further improve model accuracy.

Excluding the rigid translation, as shown in Table I, our test dataset has a mean range of the liver tumor center-of-mass motion of 13.62±6.42 mm, 6.54±4.59 mm, and 2.70±1.46 mm in the SI, AP, and LR directions, respectively. These values are comparable with previously reported intra-fractional liver tumor motion measurements via fiducial markers. Park et al. analyzed treatment CBCTs from 20 patients and reported motion ranges of 16.5 ± 5.7 mm (SI), 5.3 ± 3.1 mm (AP), and 2.8 ± 1.6 mm (LR) (Park *et al.*, 2012).

Similarly, Worm et al. measured intra-fractional motion during stereotactic body radiation therapy (SBRT) and found mean motion ranges of 16.0 mm (SI), 8.0 mm (AP), and 5.2 mm (LR) for 10 patients (Worm *et al.*, 2013). Xu et al. reported lower average motion amplitudes of 6.4 ± 5.5 mm (SI), 2.9 ± 2.8 mm (AP), and 2.1 ± 2.3 mm (LR) (Xu *et al.*, 2014), while a later study by Worm et al. using 4D-CT scans reported mean motion ranges of 11.0 mm (SI), 4.3 mm (AP), and 2.2 mm (LR) (Worm *et al.*, 2018). Our motion ranges are therefore within the range of values reported in the literature. Importantly, in addition to considering intra-fractional motion as previous studies did, our evaluation also incorporated simulated inter-fractional rigid translation of up to 10 mm to better represent realistic clinical variations. As shown in Table II, these motion scenarios underscore the value of using diffusion models for robust liver tumor localization under varying motion conditions.

The superior performance of PCD-Liver over X360 can be attributed to several key advantages of diffusion models over GNN-based architectures. Unlike most networks which operate in a single forward pass, the diffusion model in PCD-Liver operates iteratively, gradually refining predictions over multiple steps. This enables the model to capture fine-grained deformations while reducing overfitting to noise and outliers. The superiority of diffusion models over one-pass and GNN-based approaches has also been demonstrated in existing studies, where point cloud diffusion models have been shown to outperform these architectures in tasks such as point cloud registration (Chen *et al.*, 2023) and shape generation (Luo and Hu, 2021). These findings further support the advantages of using a diffusion-based framework for precise and robust liver motion tracking. Diffusion models also exhibit greater resilience to OOD data, a crucial factor in medical imaging applications. Diffusion models are designed to iteratively learn the underlying data distribution by gradually adding and removing noise, allowing them to generalize beyond their training domain and effectively denoise inputs even under high noise conditions. Evidence from Graham et al. (Graham *et al.*, 2023) demonstrates that diffusion models outperform conventional generative models and reconstruction-based approaches in detecting OOD data by leveraging an externally controlled information bottleneck rather than a fixed latent space, thus enhancing adaptability to unseen variations. Similarly, Li et al. (Li *et al.*, 2023) have shown that diffusion models achieve robust medical image translation without exposure to the test-domain data during training. Consequently, this ability contributes to PCD-Liver's stable performance under high noise conditions, reinforcing the diffusion model's advantage in handling unpredictable noise and data variations.

However, a limitation of the diffusion-based approach is time efficiency. Diffusion models involve a multi-step denoising process, where noise is gradually removed iteratively to generate high-quality predictions. While this iterative process contributes to the model's accuracy and stability, it also results in slower prediction times compared to models like X360, which is capable of near-instantaneous predictions. Real-time performance is crucial in IGRT, where quick adaptations to patient motion are necessary to ensure precise dose delivery. In general, overall temporal latency for real-time adaptive radiotherapy should be within 500 ms (Keall *et al.*, 2021). Currently, our diffusion model does not meet the speed requirements for real-time applications in IGRT as it takes around 90 s for motion estimation. To address this limitation, a promising solution is to use latent diffusion models (LDMs), which operate in a compressed latent space, thereby reducing computational load and allowing for faster generation times (Rombach *et al.*, 2022). Recent research has shown that latent diffusion models can achieve near real-time performance by decreasing the number of necessary iterations while maintaining accuracy. GameNGen (Valevski *et al.*, 2024) utilizes LDMs to simulate game frames, achieving real-time performance with 4 denoising steps and a total inference time of 50 ms per frame, enabling it to run interactively as a game engine at 20 frames per second. For point cloud latent diffusion models, it was reported that using a denoising diffusion probabilistic modeling (DDPM) (Ho *et al.*, 2020a) scheduler resulted in an inference time of 0.28 s per sample (Ji *et al.*, 2024). By employing the Denoising Diffusion Implicit Model (DDIM) (Song *et al.*, 2020) scheduler and reducing the denoising steps to 50, the runtime can be reduced to 120 ms (Chen *et al.*, 2024). Integrating latent diffusion methods into our framework can significantly improve its efficiency, bringing our diffusion-based approach closer to real-time applicability in clinical settings.

Another limitation of our approach lies in the reliance on simulated cone-beam CT (CBCT) projections for training. While the simulated projections include photon and electronic noise, they lack scatter noise, one of the major sources of signal degradation in real CBCT imaging (Endo *et al.*, 2001; Graham *et al.*, 2007). Scatter noise, arising from interactions within the patient's body, can significantly affect image quality, and its exclusion in simulated training data may limit the generalizability of our model to real-world clinical scenarios. Although Monte Carlo-based methods can simulate scatter noise accurately (Jia *et al.*, 2012), their high computational cost makes them impractical for on-the-fly simulation in training. A potential alternative could be deep learning-based approaches (Maier *et al.*, 2018; Peng *et al.*, 2019) that simulate realistic CBCT images with scatter noise in minimal inference time. Integrating these techniques could allow for more accurate simulation of clinical CBCT conditions to enhance the robustness of our model when applied to real patient data.

Future work could focus on several areas to address these limitations. First, we plan to explore the integration of latent diffusion models to enhance time efficiency while retaining accuracy. Second, incorporating realistic CBCT simulation techniques that account for scatter noise could further improve the model's generalizability and robustness in clinical applications. These advancements would further advance our model towards precise, real-time liver motion tracking and plan adaptation in radiotherapy.

## 5. Conclusion

We developed a novel framework combining a conditional point cloud diffusion model with rigid motion correction for accurate liver motion estimation from single X-ray projections. Our model demonstrated robust performance, outperforming state-of-the-art alternatives like X360 across RMSE, HD95, and COME metrics, even under highly noisy conditions. The single projection-based liver deformable motion tracking and tumor localization enables intra-treatment motion management, especially real-time image-guided adaptive radiotherapy.

## Acknowledgments

The study was supported by the US National Institutes of Health (R01 CA240808, R01 CA258987, R01 EB034691, and R01 CA280135) and Varian Medical Systems.

## Conflict of interest statement

The authors have no relevant conflicts of interest to disclose.

## Ethical statement

The datasets used in this study were retrospectively collected from an approved study at UT Southwestern Medical Center on August 31, 2023, under an umbrella IRB protocol 082013-008 (Improving radiation treatment quality and safety by retrospective data analysis). This is a retrospective analysis study and not a clinical trial. No clinical trial ID number is available. Individual patient consent was signed for the anonymized use of the imaging and treatment planning data for retrospective analysis. These studies were conducted in accordance with the principles embodied in the Declaration of Helsinki.